\pdfoutput=1

\documentclass[fleqn,usenatbib]{mnras}

\usepackage{newtxtext,newtxmath}

\usepackage[T1]{fontenc}

\DeclareRobustCommand{\VAN}[3]{#2}
\let\VANthebibliography\thebibliography
\def\thebibliography{\DeclareRobustCommand{\VAN}[3]{##3}\VANthebibliography}

\usepackage{graphicx}	
\usepackage{orcidlink}

\newcommand{\km}{\rm\thinspace km}

\newcommand{\erg}{\rm\thinspace erg}
\newcommand{\s}{\rm\thinspace s}

\newcommand{\kmps}{\hbox{$\km\s^{-1}$}}
\newcommand{\ergps}{\hbox{$\erg\s^{-1}\,$}}

\newcommand{\civ}{\ion{C}{iv}}
\newcommand{\mgii}{\ion{Mg}{ii}}

\newcommand{\oiii}{\hbox{[\ion{O}{iii}]}}
\newcommand{\hb}{H$\beta$}
\newcommand{\ha}{H$\alpha$}
\newcommand{\hg}{H$\gamma$}
\newcommand{\hd}{H$\delta$}

\title[{[O\,{\normalsize \textit{III}}] emission in BAL and non-BAL QSOs}]{[O\,{\Large {III}}] emission in $z\approx2$ quasars with and without Broad Absorption Lines}

\author[M. J. Temple et al.]{Matthew J. Temple$^{\orcidlink{0000-0001-8433-550X}}$,$^{1,2}$\thanks{E-mail: Matthew.J.Temple@durham.ac.uk}
Amy L. Rankine$^{\orcidlink{0000-0002-2091-1966}}$,$^{3}$
Manda Banerji$^{\orcidlink{0000-0002-0639-5141}}$,$^{4}$
Joseph F. Hennawi$^{\orcidlink{0000-0002-7054-4332}}$,$^{5,6}$
\newauthor
Paul\,C.\,Hewett$^{\orcidlink{0000-0002-6528-1937}}$,$^{7}$
James\,H.\,Matthews$^{\orcidlink{0000-0002-3493-7737}}$,$^{8}$
Riccardo\,Nanni$^{\orcidlink{0000-0002-2579-4789}}$,$^{5,6}$
Claudio\,Ricci$^{\orcidlink{0000-0001-5231-2645}\,2,9}$
and Gordon\,T.\,Richards$^{\orcidlink{0000-0002-1061-1804}\,10}$
\\ 
$^{1}$Centre for Extragalactic Astronomy, Department of Physics, Durham University, South Road, Durham DH1 3LE, UK\\
$^{2}$Instituto de Estudios Astrof\'{\i}sicos, Universidad Diego Portales, Av. Ej\'ercito Libertador 441, Santiago 8370191, Chile\\
$^{3}$Institute for Astronomy, University of Edinburgh, Royal Observatory, Blackford Hill, Edinburgh EH9 3HJ, UK\\
$^{4}$School of Physics \& Astronomy, University of Southampton, Southampton, SO17 1BJ, UK\\
$^{5}$Leiden Observatory, Leiden University, PO Box 9513, NL-2300 RA, Leiden, the Netherlands\\
$^{6}$Department of Physics, University of California, Santa Barbara, CA 93106, USA\\
$^{7}$Institute of Astronomy, University of Cambridge, Madingley Road, Cambridge CB3 0HA, UK\\
$^{8}$Department of Physics, Astrophysics, University of Oxford, Denys Wilkinson Building, Keble Road, Oxford, OX1 3RH, UK\\
$^9$Kavli Institute for Astronomy and Astrophysics, Peking University, Beijing 100871, China\\
$^{10}$Department of Physics, Drexel University, 32 S. 32nd Street, Philadelphia, PA 19104, USA
}

\date{Accepted 2024 June 17. Received 2024 June 10; in original form 2023 September 29}

\pubyear{2024}

\begin{document}
\label{firstpage}
\pagerange{\pageref{firstpage}--\pageref{lastpage}}
\maketitle

\begin{abstract}
Understanding the links between different phases of  outflows from active galactic nuclei is a key goal in extragalactic astrophysics.
Here we compare \oiii\,$\lambda\lambda$4960,5008 outflow signatures 
{in quasars with and without Broad Absorption Lines (BALs),}
aiming to test how the broad absorption troughs seen in the rest-frame ultraviolet are linked to the narrow line region outflows seen in the rest-frame optical.
We present new near-infrared spectra from Magellan/FIRE which cover \oiii\ in 12 quasars with $2.1<z<2.3$, selected to have strong outflow signatures in \civ$\,\lambda$1550. 
Combining with data from the literature,  we build a sample of  {73 BAL, 115 miniBAL  and 125 non-BAL} QSOs with $1.5<z<2.6$.
{The strength and velocity width of}
\oiii\ correlate strongly with the \civ\ emission properties, but no significant difference is seen in the \oiii\ 
{emission-line properties}
between the BALs, non-BALs and miniBALs once the dependence on \civ\ emission is taken into account.
A weak correlation is observed between the velocities of  \civ\ BALs and \oiii\ emission, which is accounted for by the fact that both outflow signatures correlate with the underlying \civ\ emission properties.
Our results add to the growing evidence that BALs and non-BALs are drawn from the same parent population and are consistent with a scenario wherein BAL troughs are intermittent tracers of persistent quasar outflows, with a part of such outflow becoming optically thick along our line-of-sight for sporadic periods of time within which BALs are observed.
\end{abstract}

\begin{keywords}
quasars:emission lines -- quasars:absorption lines
\end{keywords}



\section{Introduction}

Quasar-driven outflows are widely invoked in galaxy formation models in order to reproduce the observed properties of massive galaxies \citep[e.g.][]{1998A&A...331L...1S, 2005MNRAS.361..776S, 2006MNRAS.370..645B, Harrison17}. Luminous quasars are powerful sources of radiation, and if outflows from the active nucleus can propagate to galaxy scales then the energy in such an outflow would be enough to disrupt the interstellar medium of the host galaxy, preventing star formation and providing an explanation for the observed `co-evolution' between supermassive black holes and their hosts \citep[][]{Magorrian98, KH13}.

High-velocity outflows have long been known to exist in many luminous quasars \citep{2015ARA&A..53..115K}. Outflow velocities of many thousands of \kmps\ are common and are believed to originate from material in a wide-angle outflowing disc-wind \citep{1995ApJ...451..498M, 2019A&A...630A..94G}. Such disc-winds are now used to explain the blue-asymmetric profiles of the high-ionization \civ\,$\lambda$1550 emission line \citep{Richards11, 2020MNRAS.492.5540M, Matthews23, 2023MNRAS.524.5497S, Temple23, 2023arXiv230902491G}. Up to 50 per cent of quasars exhibit strong, blueshifted absorption due to outflowing material present directly along the line of sight \citep{1991ApJ...373...23W, 2002ApJS..141..267H, 2011MNRAS.410..860A, Rankine20, 2023ApJ...952...44B}, and the outflow speeds in such `broad absorption line' (BAL) quasars can exceed 50,000\,\kmps \citep{2019A&A...630A.111B, 2020ApJ...896..151R,  2022ApJ...939L..24R}.

Opinions differ as to whether the observed presence of BAL troughs represents a particular evolutionary phase in the quasar fuelling and outflow life-cycle, or a special viewing angle, or if instead BALs are a short, intermittent phase which all quasars will go through stochastically.
To address this question, it is informative to consider complementary tracers of Active Galactic Nuclei (AGN) winds which probe different phases and different locations in the outflow \citep{2017A&A...601A.143F}.
One popular probe of ionized gas kinematics in distant galaxies is the \oiii\,$\lambda\lambda$4960,5008\footnote{
Vacuum wavelengths are used throughout this paper: the \oiii\ doublet has $\lambda\lambda=5006.8$,4958.9 in air and $\lambda\lambda=5008.2$,4960.3 in vacuum.}
emission doublet, which is usually inferred to originate in the `narrow line region' on scales of up to
$\sim$kilo-parsecs
\citep{Baskin05, 2018MNRAS.477.4615D}.
Over recent years, many authors have studied the \oiii\ emission properties across different sub-classes of AGN, finding that more luminous quasars generally show weaker \oiii\ emission which is broader and often blueshifted, suggesting that \oiii\ in luminous quasars is tracing AGN outflows in low-density ionized gas on scales which may be important for host galaxy feedback
\citep[e.g.][]{2009A&A...495...83M, Liu13, 2014Natur.513..210S, Zakamska14, Harrison14, Harrison16, Shen16, 2017A&A...598A.122B, 2017FrASS...4...16M,  Temple19, 2020A&A...642A.147K, 2020A&A...634A.116V}.

However,  only a handful of studies have attempted to link quasar outflow signatures that are potentially probing different physical scales 
(e.g.\ \citealt{2000ApJ...545...63E, 2002ApJ...576L...9Z, 2019A&A...630A.111B, 2020MNRAS.495..305X, 2020ApJ...893...95Y}).
A holistic understanding of outflow properties probed using  different diagnostics is the only way to fully understand the effect AGN feedback has on galaxy formation.
Using a sample of $\sim$200 luminous non-BAL quasars, \citet{Coatman19} found a correlation between the outflow kinematics of the rest-frame ultraviolet \civ\,$\lambda$1550 emission produced on $\lesssim$ parsec scales \citep{2023arXiv230702914F, 2023A&A...672A..45H, 2024ApJS..272...26S} and the kinematics of the rest-frame optical \oiii\,$\lambda\lambda$4960,5008 emission believed to originate on much larger scales.
This result has also been found in smaller samples selected via X-rays \citep{2020A&A...644A.175V} and via their high luminosities in either the \textit{WISE} mid-infrared bands \citep{2018A&A...617A..81V} or in  optical photometry \citep{2023A&A...669A..83D}.
Most importantly, these correlations are still seen even when the dependence of both \civ\ and \oiii\ on the quasar luminosity has been taken into account. 
These results are consistent with a scenario wherein nuclear outflows traced by the \civ\ emission are capable of propagating to galaxy-wide scales, where they would be able to return a significant amount of energy to the interstellar medium of their host galaxies. 

To better understand how BAL outflows are linked to emission-line blueshifts, \citet{Rankine20} measured the {\civ} emission line parameters and BAL properties for $\simeq$140\,000 quasar spectra from the Sloan Digital Sky Survey (SDSS) DR14 quasar catalogue \citep{Paris18}. Spectrum reconstructions based on an independent component analysis (ICA) of a sample of non-BAL quasars allowed for the intrinsic {\civ} emission of both the BAL and non-BAL quasars to be robustly reconstructed and measured, even in the presence of extensive absorption. \citet{Rankine20} found that the {\civ} emission properties of the BAL and non-BAL quasar populations were extremely similar, suggesting that BAL and non-BAL quasars represent different views of the same underlying quasar population. Additionally, BAL trough properties such as the maximum and minimum absorption velocities and the BALnicity index (a measure of the amount of absorption; see Section~\ref{sec:data:SDSS} for the definition) were found to strongly correlate with {\civ} emission line properties.

To further test the hypothesis put forward by \citet{Rankine20}, viz. that BALs and non-BALs are drawn from the same parent population,
in this paper we investigate the narrow line region \oiii\ emission in a large sample of 73 BAL quasars and compare with the non-BAL population to test whether BALs 
{show evidence for being in a special evolutionary phase.}
More precisely, in a model where the BAL outflows represent a specific phase in the quasar fuelling/outflow cycle,
the extended narrow line region emission would be expected to differ significantly between the BAL and non-BAL quasars \citep{1997ApJ...476...40T}. 
The impact of the energetic BAL flows would reduce the emission from the static narrow line region although broader, more blueshifted \oiii\  emission may be seen \citep[e.g.][]{Zakamska16}.
At a given \civ\ emission-line blueshift and equivalent width (EW), \oiii\ in the BAL quasars would have lower EW and be more blueshifted compared to the non-BALs.
Alternatively, if  BALs are stochastic phenomena which may appear intermittently for short periods of time while any galaxy is in a luminous quasar phase, then their \oiii\ properties should be very similar to non-BALs. 

Previous near-infrared observations of $z>1$ quasars have mostly focused on non-BAL quasars \citep[e.g.][]{Coatman19}, and so in Section~\ref{sec:data} we present 12 new Magellan/FIRE spectra which were specifically targeted to observe the rest-frame optical \oiii\ in BAL quasars. 
We combine with archival data from the literature to build a sample of 73 BAL, 115 miniBAL and 125 non-BAL quasars, 
as defined in Section~\ref{sec:data:SDSS}.
In Section~\ref{sec:methods} we then measure the \oiii\ strength and kinematics in our sample of BAL quasars, and  compare the \oiii\ outflow signatures seen in the BAL, miniBAL and non-BAL populations. 
We discuss our findings in Section~\ref{sec:discuss} and summarise our conclusions in Section~\ref{sec:conclude}.

\section{Sample and Data}
\label{sec:data}

\begin{figure}
    \centering
    \includegraphics[width=\columnwidth]{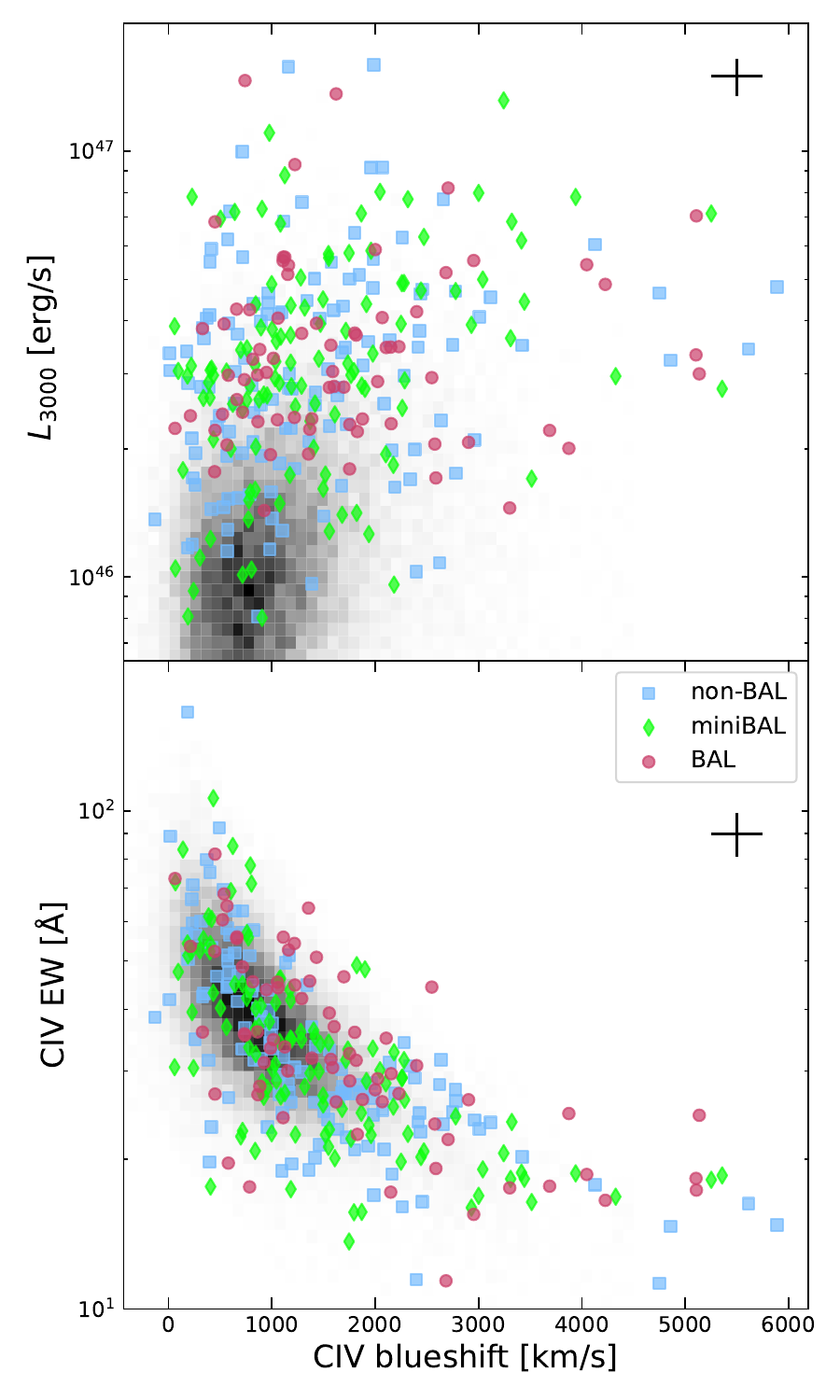}
    \caption{Distribution of our near-infrared sample as a function of rest-frame  3000\,\AA\ continuum luminosity ($L_{3000}$) and  the \civ\ emission-line blueshift and EW.
    The typical uncertainty associated with each individual data point is shown by the black crosses in the upper right of each panel.
    In gray we show the distribution of our parent sample of SDSS  quasars with $1.5<z<2.6$ and signal-to-noise ratio $>10$. Our sample of objects with near-infrared data covers the mode of the SDSS population in emission-line space, while extending to brighter luminosities and larger \civ\ blueshifts to probe faster outflow signatures.
    The median ultraviolet continuum luminosity of each of our BAL, miniBAL and non-BAL subsamples is $L_{3000}=3\times10^{46}$\,\ergps, corresponding to $L_\textrm{bol}\approx1.5\times10^{47}$\,\ergps.
    }
    \label{fig:C4_samples}
\end{figure}

\begin{figure*}
    \centering
    \includegraphics[width=\columnwidth]{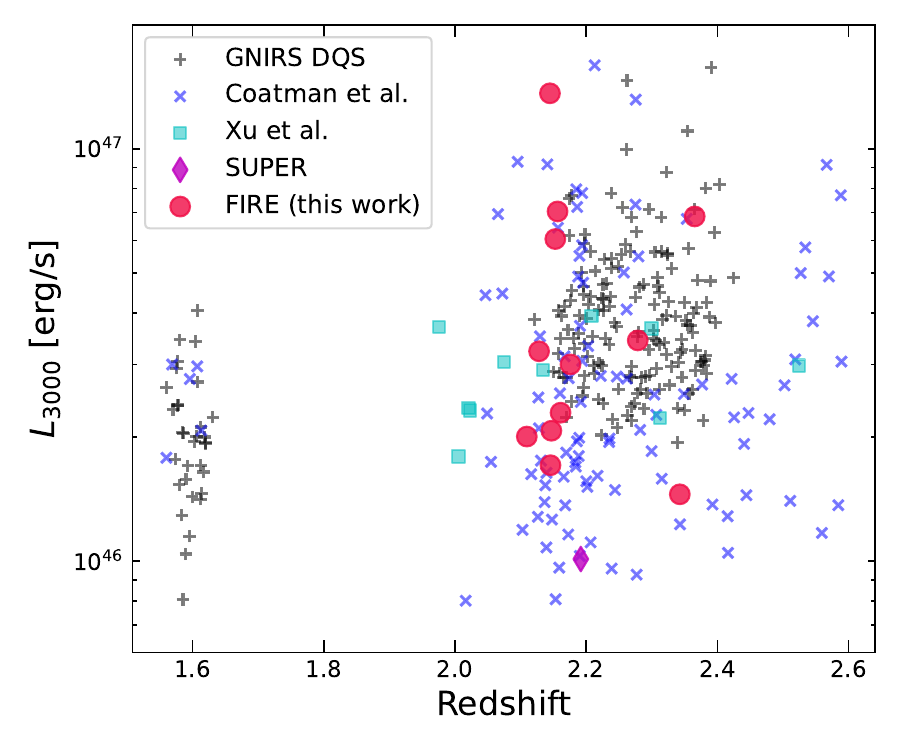}
    \includegraphics[width=\columnwidth]{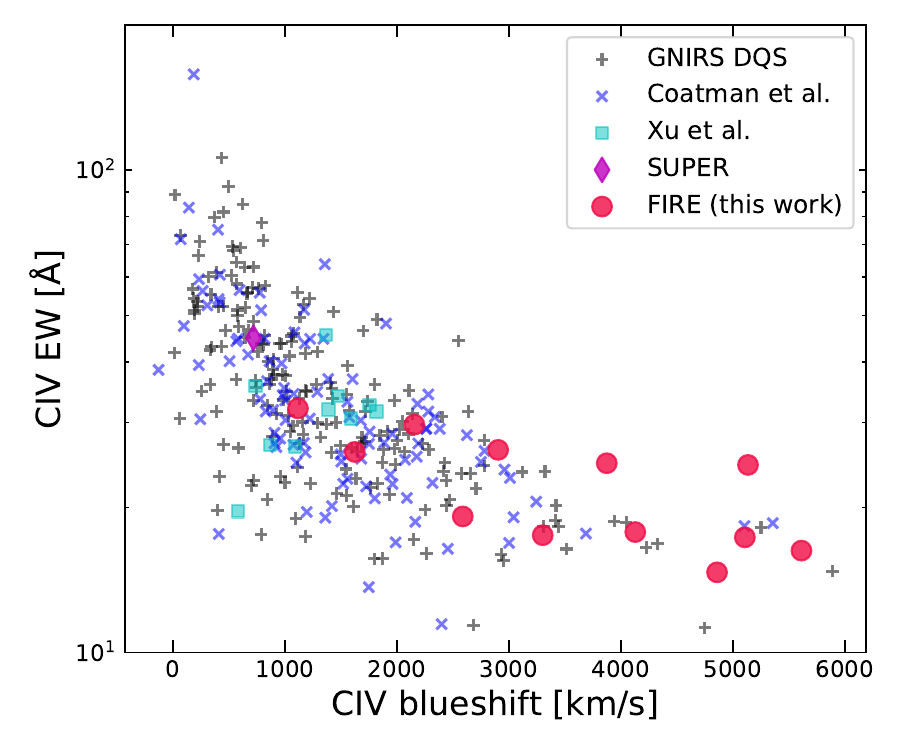}
    \caption{Distributions of redshift, luminosity, and \civ\ emission properties for the different near-infrared samples compiled in this work.
    Our new FIRE observations prioritised targets with larger \civ\ blueshifts, as such objects are rare in existing samples.
    }
    \label{fig:zlum}
\end{figure*}

To investigate the link between the outflows seen in the rest-frame ultraviolet and rest-frame optical wavebands, we compile a sample of $1.5<z<2.6$ quasars with both SDSS observed-frame optical spectra (subsection~\ref{sec:data:SDSS}) and near-infrared spectra from various sources (subsections~\ref{sec:data:FIRE} and \ref{sec:data:existingNIR}).
To allow identification of broad absorption features up to 25\,000\kmps\ bluewards of \civ, we require redshift $z > 1.56$ for objects with spectra from BOSS and $z>1.67$ for objects with data from the original SDSS spectrograph (i.e.\ observations before the `SDSS MJD' of~55000). 
The $z<2.6$ criterion allows us to check for low-ionization \mgii\ broad absorption troughs in our final sample; no such `LoBALs' are found. To ensure reliable detection of absorption troughs in the SDSS spectra we require the average signal-to-noise ratio per pixel to be $\ge10$, as the BAL fraction has been found to depend only weakly on the signal-to-noise ratio above this threshold (fig.\ 2 of \citealt{Rankine20}, see also fig. 4 of \citealt{2009ApJ...692..758G}).
To ensure coverage of \oiii, \hb\ and \ha\ in the infrared \textit{JHK} bands, we require either 
$1.56<z<1.65$ or $1.95<z<2.60$, to avoid emission lines falling in the regions of low atmospheric transparency between the \textit{JHK} bands.

We measure the  rest-frame ultraviolet monochromatic luminosity $\lambda L_\lambda$ at $\lambda=3000$\,\AA\ (hereafter $L_{3000}$) by fitting a model spectral energy distribution \citep{THB21} to the \textit{griz} SDSS photometric data.
Our final sample spans $8 \times 10^{45}<L_{3000}<2\times10^{47}\,\ergps$.
The bolometric correction for each object is likely in the range $L_\textrm{bol}/L_{3000} \approx3$-10 \citep[][fig.~6]{Temple23},  so all of the objects  in our sample lie well above the 
 $L_\textrm{bol}\gtrsim3\times10^{45}\,\ergps$ threshold which was suggested by \citet{Zakamska16} as necessary for \oiii\ winds to contribute to quasar feedback.

Our compilation results in a total of 313 unique SDSS quasars with optical and near-infrared spectra covering the rest-frame 1400--2800 and 4800--6600\,\AA\ wavelength ranges, as shown in Figs.~\ref{fig:C4_samples} and \ref{fig:zlum}.
The majority of our sample (276/313)
have $1.95<z<2.6$; only 37
quasars have $1.56<z<1.65$.
We have verified that the results and conclusions of this paper would not change if we restricted our sample to $2<z<2.5$ and $10^{46}<L_{3000}<10^{47}\,\ergps$.
From our sample of 313 quasars,
125 objects show no \civ\ absorption exceeding $\approx 450$\,\kmps\ in the rest-frame ultraviolet (`non-BALs'),
115 show mild absorption with trough widths $>450$\,\kmps\ which does not meet the strict definition of a BAL (`miniBAL' quasars), 
while 73 sources are \textit{bona fide} BALQSOs according to the definition of \citet{1991ApJ...373...23W}.
All of our BALQSOs are so-called `HiBAL' systems which show no evidence for low-ionization absorption troughs.

We now describe in detail the data sets used, including the SDSS optical data (subsection~\ref{sec:data:SDSS}), 12 new near-infrared observations from the FIRE spectrograph (\ref{sec:data:FIRE}) and existing near-infrared spectra from the literature (\ref{sec:data:existingNIR}).
In Section~\ref{sec:methods} we then describe the methods used to analyse these spectra.

\subsection{SDSS spectra and \ion{C}{IV} measurements}
\label{sec:data:SDSS}

We use the same \civ\ emission-line information as described in section~2.1 of \citet{Temple23}. In brief, we start with all quasars from the 16th and 17th data releases of SDSS \citep{2020ApJS..250....8L, 2022ApJS..259...35A}. 
For this work, we consider only quasar spectra with mean signal-to-noise ratio  (per 69\kmps\ pixel) $\ge10$
over the rest-frame interval 1700-2200\,\AA. 
Each spectrum is reconstructed using the independent component analysis (ICA) scheme described by \citet{Rankine20}.
A linear combination of ten spectral ICA components is used to model the data, using an iterative routine to mask absorption features. This approach allows the underlying emission-line properties to be inferred consistently in both the BAL and non-BAL quasars. 
Examples of the SDSS data and corresponding ICA reconstructions are shown in Appendix~\ref{sec:app:SDSSspectra}.
We keep only ICA reconstructions with reduced-$\chi^2<2$.
The \civ\ emission-line blueshift is measured as the Doppler shift of the median continuum-subtracted line flux, assuming a rest-frame wavelength of 1549.48\,\AA.

The uncertainty on the \civ\ blueshift is dominated by the systemic redshift uncertainty, which is  $\lesssim250$\,\kmps\ for our $z\approx2$ quasars \citep{2010MNRAS.405.2302H}.
The uncertainties associated with our \civ\ emission-line EWs are dominated by the time variability of individual quasars, as for example investigated by
\citet{2020ApJ...899...96R} using the SDSS-RM sample of high-cadence repeat spectroscopic observations.
From fig.~9 of \citet{2020ApJ...899...96R} we see that the \civ\ EW can vary by up to 20 per cent as a function of observation epoch.

BAL quasars are identified via the BALnicity Index \citep[BI; ][]{1991ApJ...373...23W} and the Absorption Index \citep[AI; ][]{2002ApJS..141..267H}. The former is defined as
\begin{equation}
    \text{BI} = \int_{3000}^{25000} \left(1-\frac{f(V)}{0.9}\right)C\ \text{d}V,
    \label{eq:BI}
\end{equation}
with $C=1$ when $f(V)<0.9$ contiguously for at least 2000\kmps. $f(V)<0.9$ is the continuum-normalised spectrum; in this case normalised by the reconstruction. The lower integration limit of 3000{\kmps} is set to remove any contribution of strong `associated absorbers' while the upper 25\,000{\kmps} limit avoids confusion with absorption of the \ion{Si}{IV}\,$\lambda$1397 ion.
The absorption index, on the other hand, includes absorption below 3000{\kmps} and requires $f(V)<0.9$ contiguously for at least only 450\kmps:
\begin{equation}
    \text{AI} = \int_{0}^{25000} \left(1-\frac{f(V)}{0.9}\right)C\ \text{d}V.
    \label{eq:AI}
\end{equation}
There exists a bimodal distribution of $\log$(AI) first observed by \citet{2008MNRAS.386.1426K}: one population of quasars have both AI$>$0 and BI$>$0, where, for the majority, the AI and BI are measuring the same absorption trough(s). The second population have AI$>$0 but BI$=$0 due to the presence of only narrow troughs, or broad troughs where a significant fraction of the absorption occurs below 3000\kmps.
Here we use `miniBAL' to refer to the second AI$>$0 population with BI$=$0 i.e. quasars which have absorption troughs wider than 450\,\kmps\ without meeting the \citet{1991ApJ...373...23W} definition of a BAL.
Note that many of these `miniBAL' systems are in fact narrow \civ\ doublet absorption features with $v_\textrm{doublet}=499$\,\kmps (see fig.~18 of \citealt{Rankine20}), or `line-locked' triplet systems which provide unambiguous evidence of radiation line driving playing an important role in AGN disc-winds \citep{2014MNRAS.445..359B, 2023ApJ...945..110L}.

\subsection{New Magellan/FIRE spectra}
\label{sec:data:FIRE}

We were awarded two nights on Magellan with FIRE \citep{2013PASP..125..270S} to obtain near-infrared spectra of quasars with unusual \civ\ emission line properties (CN2020B-4; PI: Temple). 
Targets were selected from the SDSS DR14 quasar catalogue \citep{Paris18, Rankine20} to have redshifts $2.1<z<2.4$, ensuring good observability of \oiii\ and \hb\ in the \textit{H} band and \ha\ in the \textit{K} band.
We further required matches to VIKING \citep{Edge13}, VHS \citep{McMahon13}, UKIDSS-LAS \citep{Lawrence07} or 2MASS \citep{2MASS} photometry with either $H<17.1$ or $(J+K)/2 < 17.1$ to ensure good signal-to-noise ratio in the resulting near-infrared spectra.
Targets were then prioritised based on their \civ\ properties: objects with large \civ\ blueshifts  were preferred, as these are under-represented in existing samples.
Targets were observed on the nights of 2022 January 02 and 03 using FIRE in echelle mode with the 0.6\,arcsec slit, which delivers $R = 6000$ spectra across  0.82--2.51\,\micron.
Seeing was in the range 0.5 to 0.9\,arcsec with dark moon and no clouds.
12 science targets were observed, including seven BALs and three non-BALs with CIV blueshift $>2000\kmps$ and two filler targets (one BAL and one non-BAL) with shorter exposures, as summarised in Table~\ref{tab:log}.

\begin{table*}
    \centering
    \caption{Quasars with new near-infrared spectra obtained from Magellan/FIRE, as described in Section~\ref{sec:data:FIRE}.}
    \label{tab:log}
    \begin{tabular}{l|c|c|c|c|c|c|c|c|c|c}
        \hline
        SDSS name & Classification & \ion{C}{IV} EW & \ion{C}{IV} blueshift & $z_\textrm{uv}$ & \textit{i}$_\textrm{SDSS}$ & \textit{J} & \textit{H} & \textit{K} & \textit{JHK} source & $t_\textrm{exp}$\\
            &   &   \AA\ & \kmps\  &  & AB & Vega & Vega & Vega &  & min\\
        \hline
J010754.95-095744.2 & BAL & 19.1 & 2587 & 2.146 & 18.55 & 17.37 & -- & 16.14 & VHS & 90\\ 
J011934.27+052629.7 & BAL & 17.5 & 3302 & 2.343 & 18.93 & 17.48 & 16.77 & 15.99 & UKIDSS & 80\\ 
J014725.50-101438.9 & non-BAL & 17.8 & 4128 & 2.153 & 17.09 & 16.12 & 15.36 & 14.78 & 2MASS & 30\\ 
J022943.90-003458.0 & BAL & 26.3 & 2903 & 2.147 & 18.19 & 17.01 & 16.51 & 15.68 & VHS & 60\\ 
J025345.28-080353.3 & BAL & 24.7 & 3874 & 2.109 & 18.41 & 17.26 & -- & 15.96 & VHS & 90\\ 
J094748.06+193920.0 & non-BAL & 16.3 & 5613 & 2.279 & 17.78 & 16.98 & 16.46 & 15.78 & 2MASS & 60\\ 
J100711.80+053208.8 & BAL & 26.1 & 1622 & 2.145 & 16.18 & 15.16 & 14.62 & 13.85 & UKIDSS & 20\\ 
J103546.02+110546.4 & non-BAL & 32.1 & 1115 & 2.365 & 17.26 & 15.77 & 15.10 & 14.23 & UKIDSS & 20\\ 
J120508.10+013455.8 & BAL & 29.7 & 2155 & 2.161 & 18.21 & 16.64 & 16.06 & 15.30 & VIKING & 40\\ 
J120550.19+020131.5 & BAL & 17.3 & 5108 & 2.156 & 16.94 & 16.02 & 15.61 & 15.00 & VIKING & 40\\ 
J121328.78-025617.8 & BAL & 24.5 & 5136 & 2.176 & 17.84 & 16.95 & 16.42 & 15.69 & VIKING & 60\\ 
J123647.13+185311.3 & non-BAL & 14.7 & 4859 & 2.128 & 17.64 & 16.69 & -- & 15.32 & 2MASS & 40\\ 
\hline
    \end{tabular}
\end{table*}

\begin{figure*}
    \centering
    \includegraphics[clip=true, trim={10 35 5 10}, width=\columnwidth]{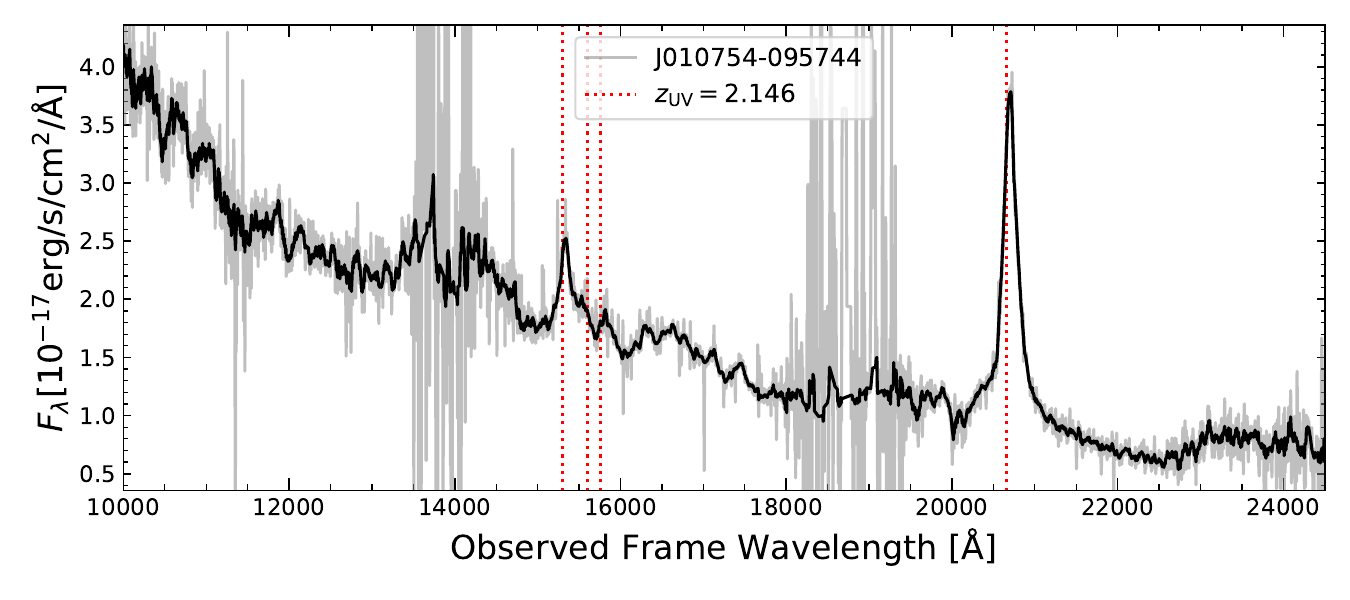}
    \includegraphics[clip=true, trim={10 35 10 10}, width=\columnwidth]{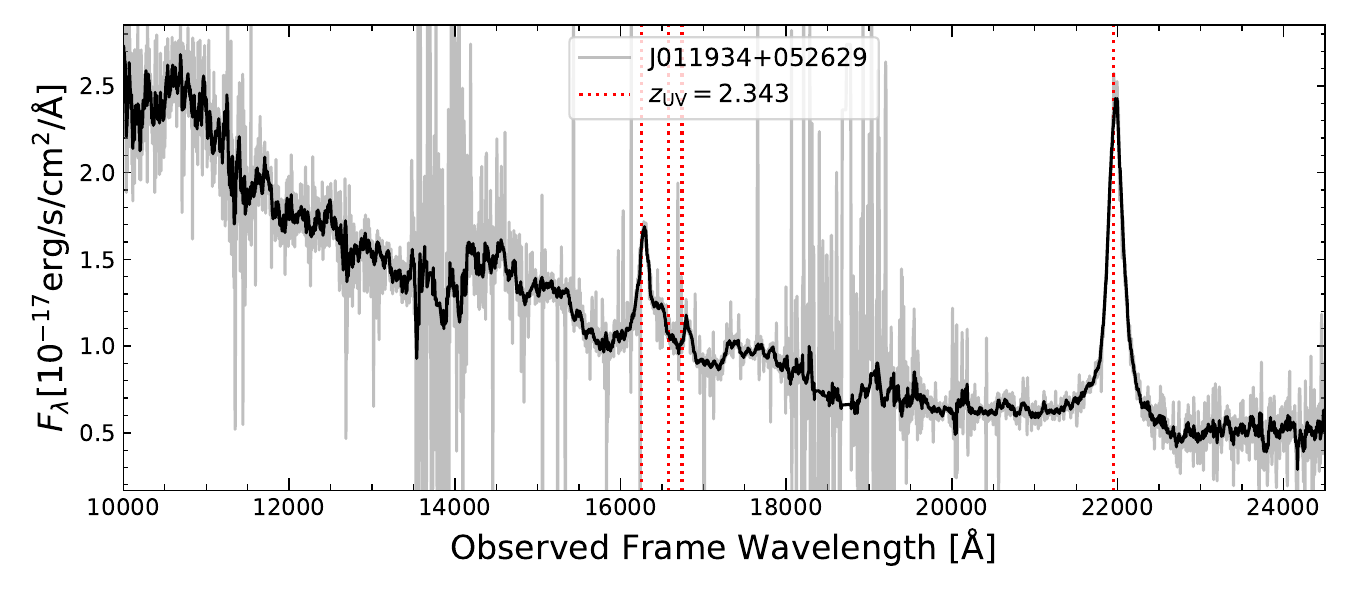}
    \includegraphics[clip=true, trim={10 35 5 10}, width=\columnwidth]{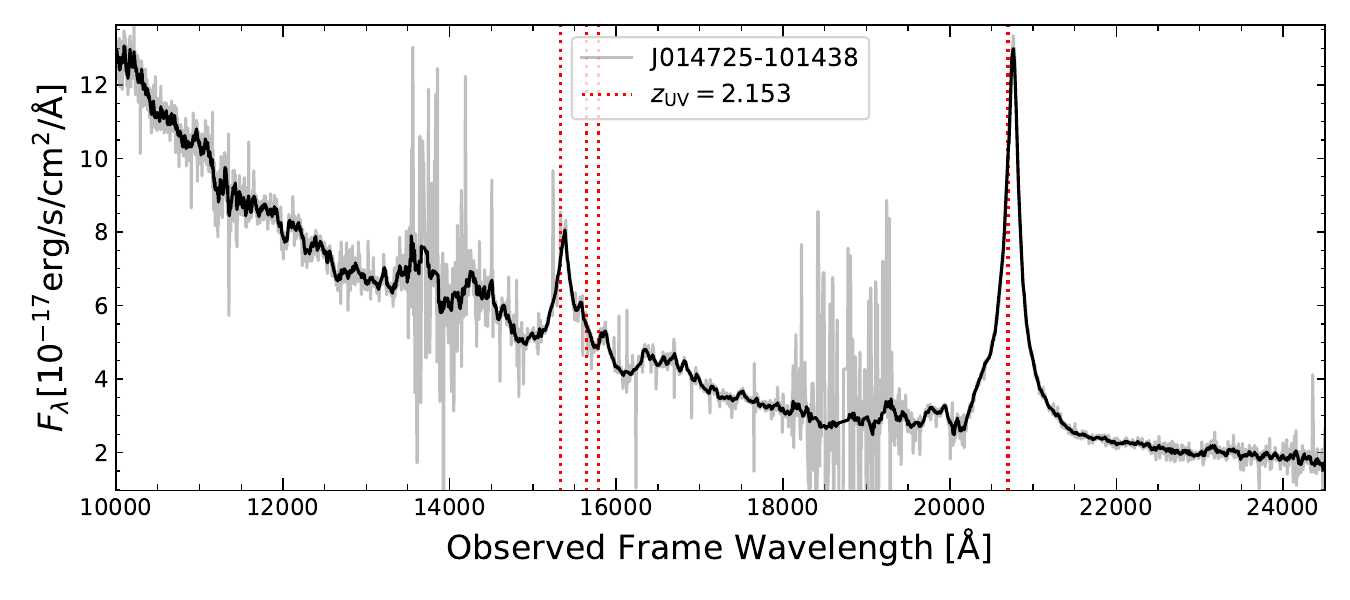}
    \includegraphics[clip=true, trim={10 35 10 10}, width=\columnwidth]{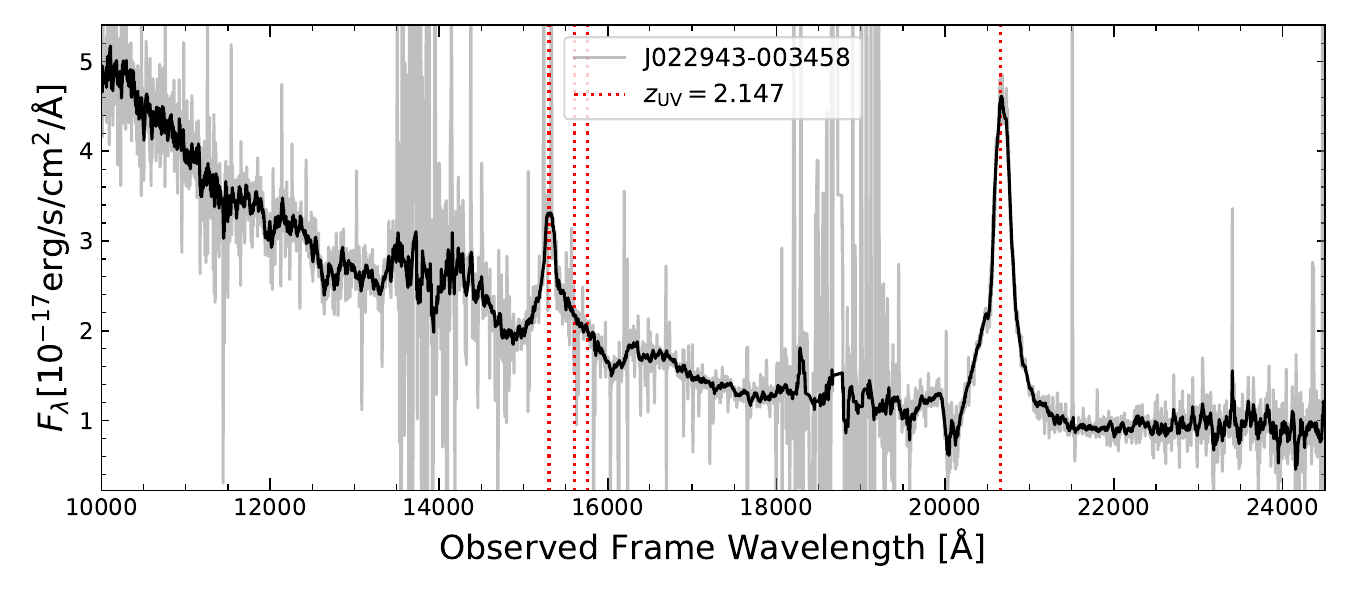}
    \includegraphics[clip=true, trim={10 35 5 10}, width=\columnwidth]{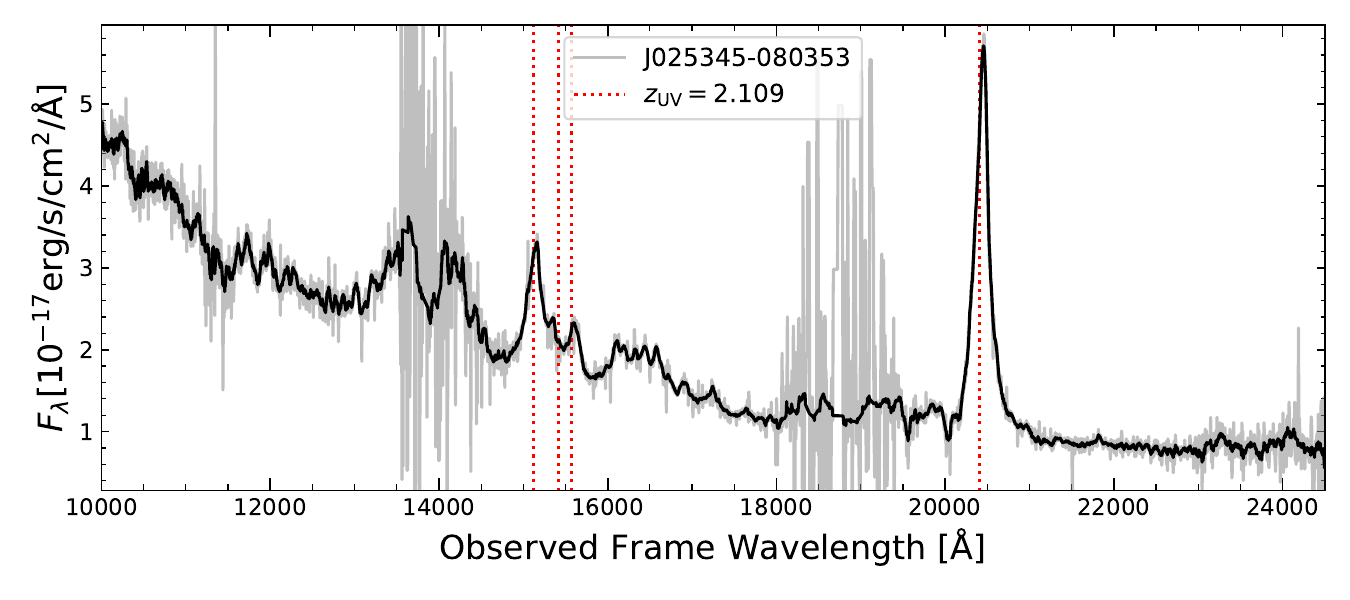}
    \includegraphics[clip=true, trim={10 35 10 10}, width=\columnwidth]{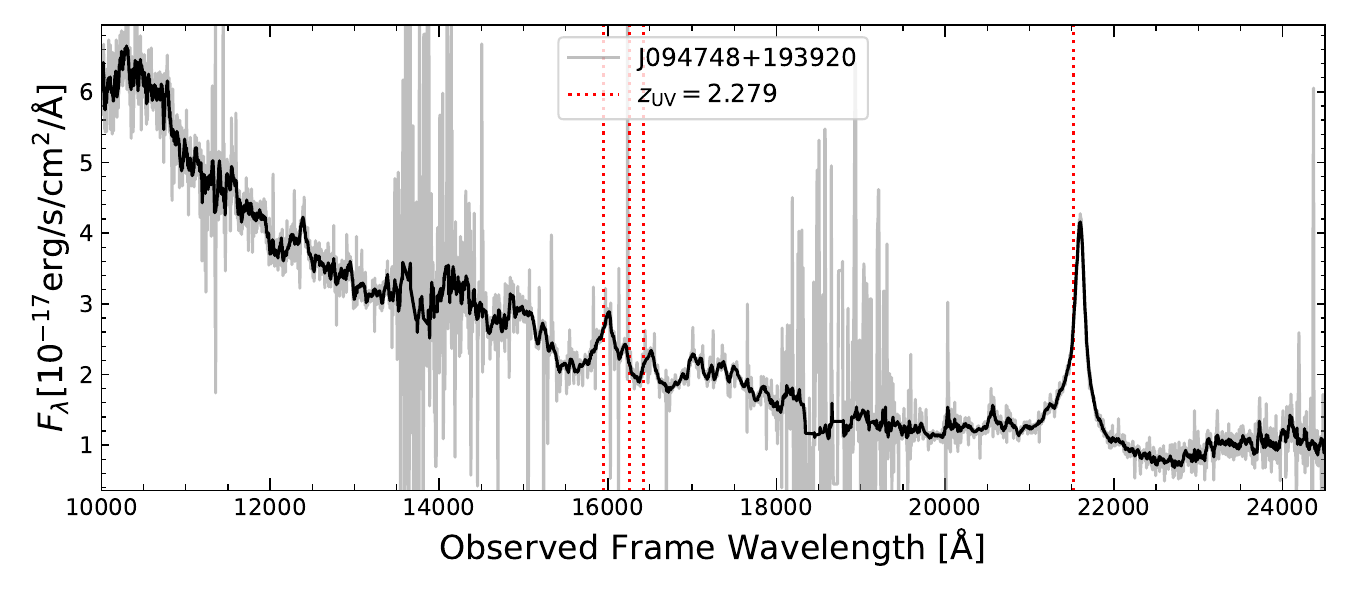}
    \includegraphics[clip=true, trim={10 35 5 10}, width=\columnwidth]{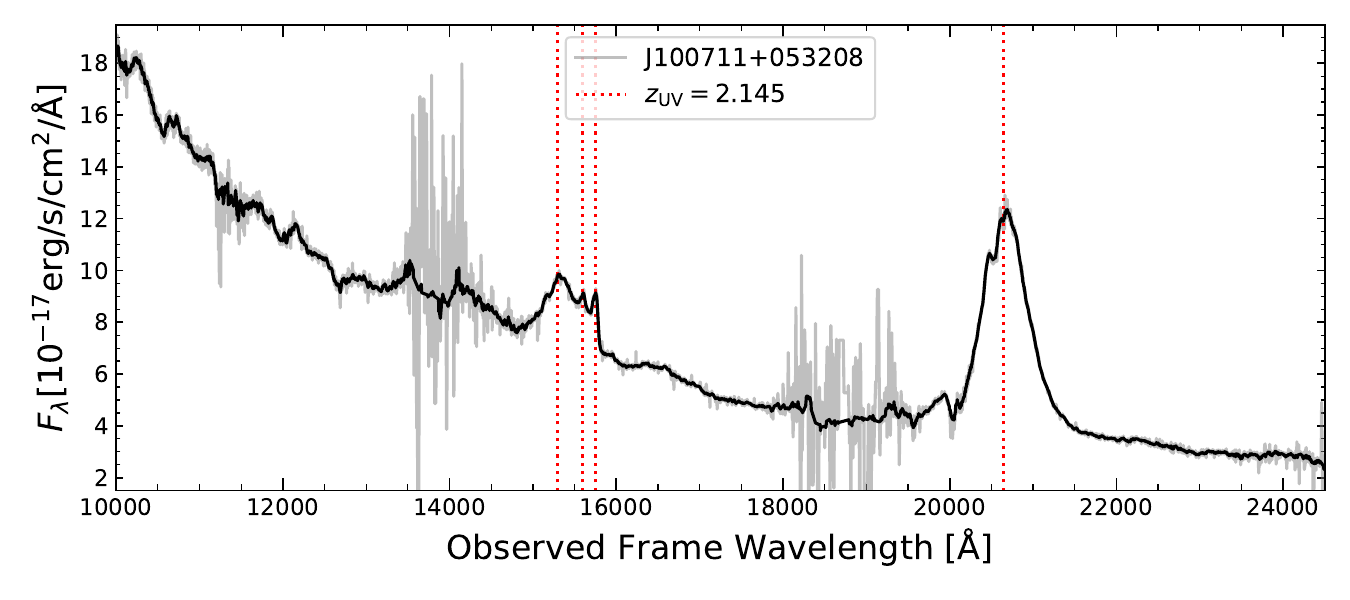}
    \includegraphics[clip=true, trim={10 35 10 10}, width=\columnwidth]{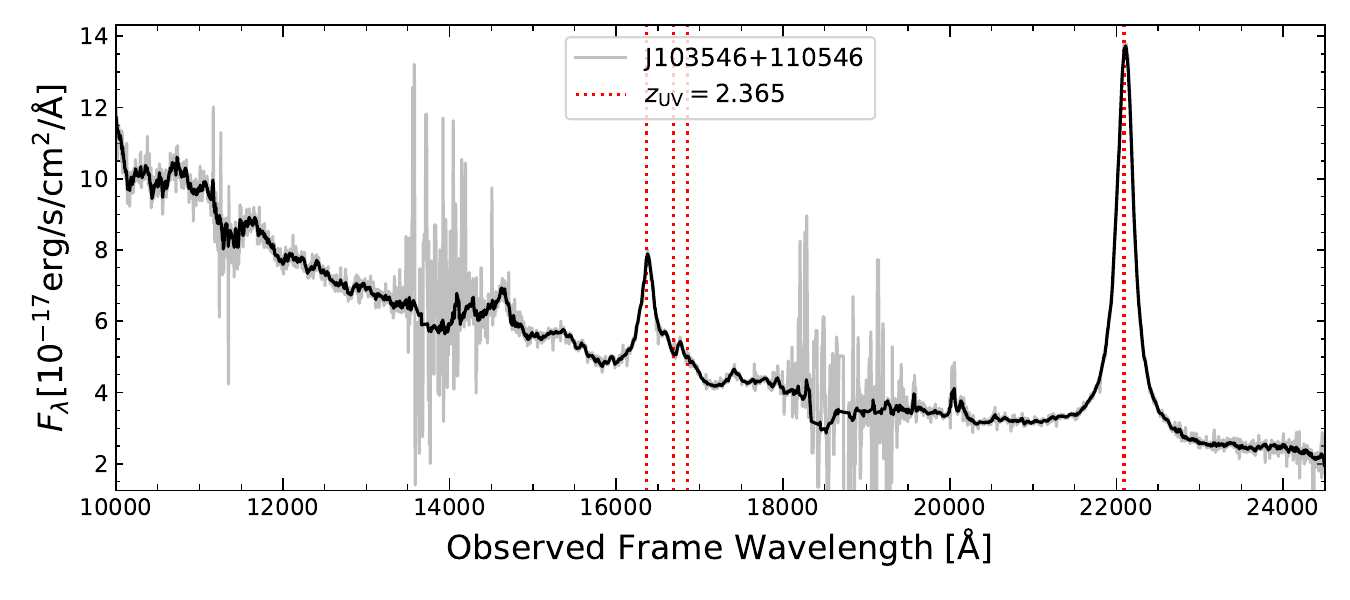}
    \includegraphics[clip=true, trim={10 35 5 10}, width=\columnwidth]{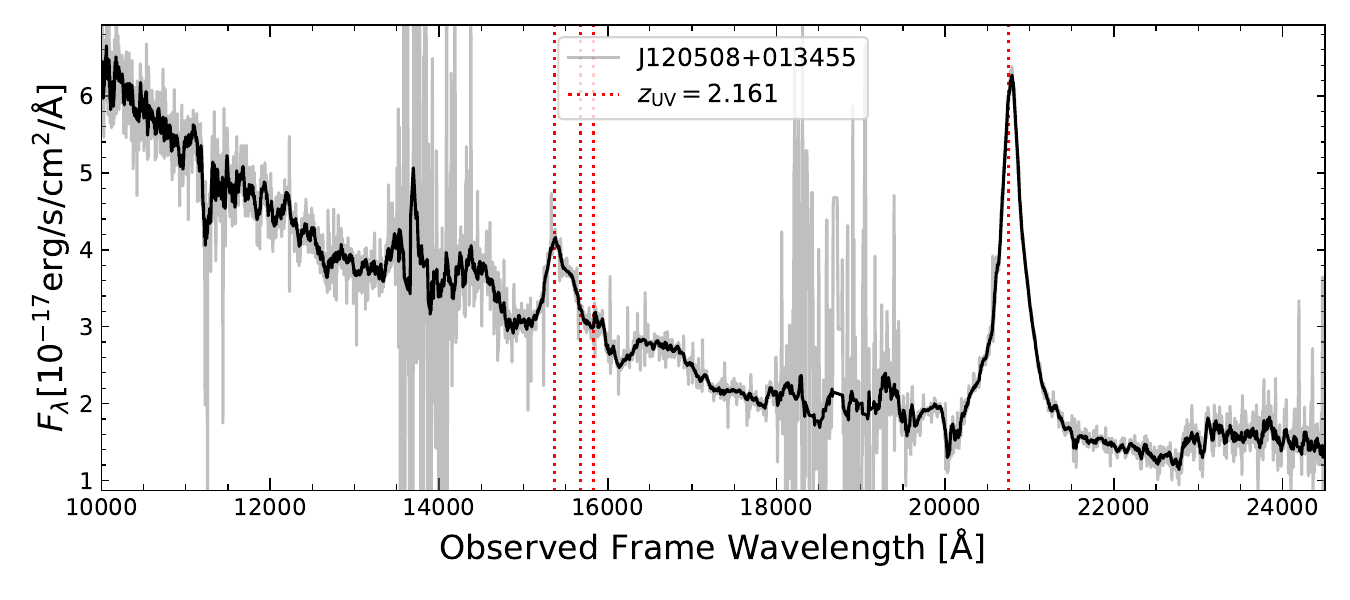}
    \includegraphics[clip=true, trim={10 35 10 10}, width=\columnwidth]{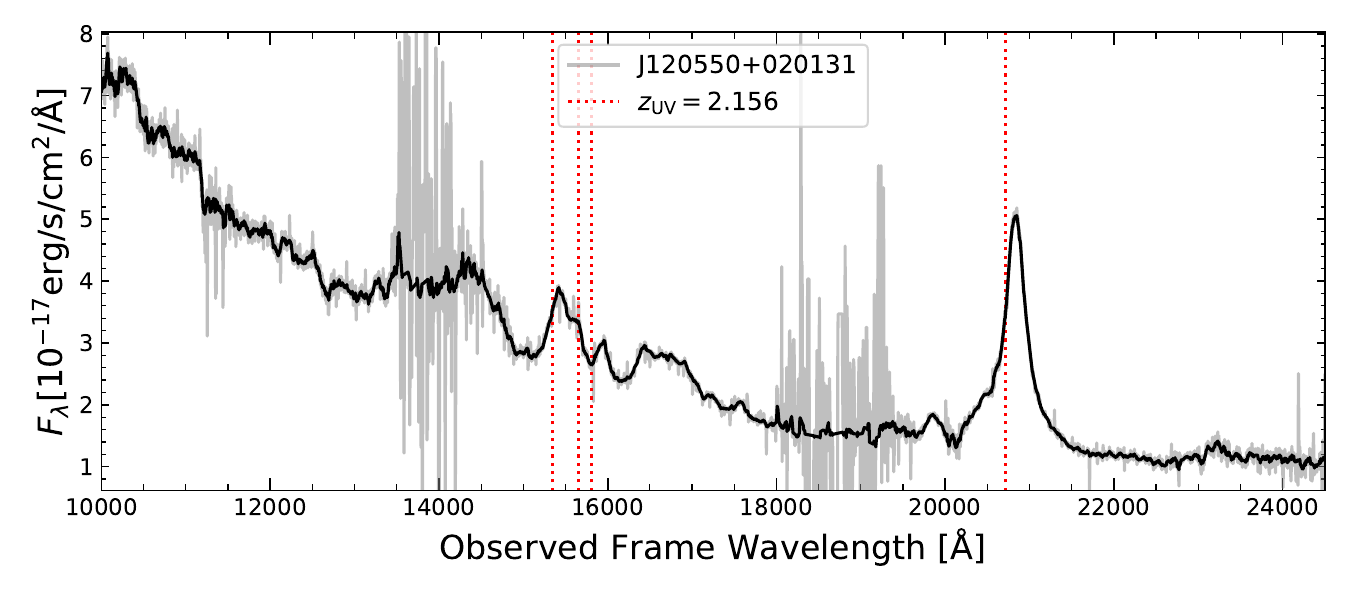}
    \includegraphics[clip=true, trim={10 0 5 10}, width=\columnwidth]{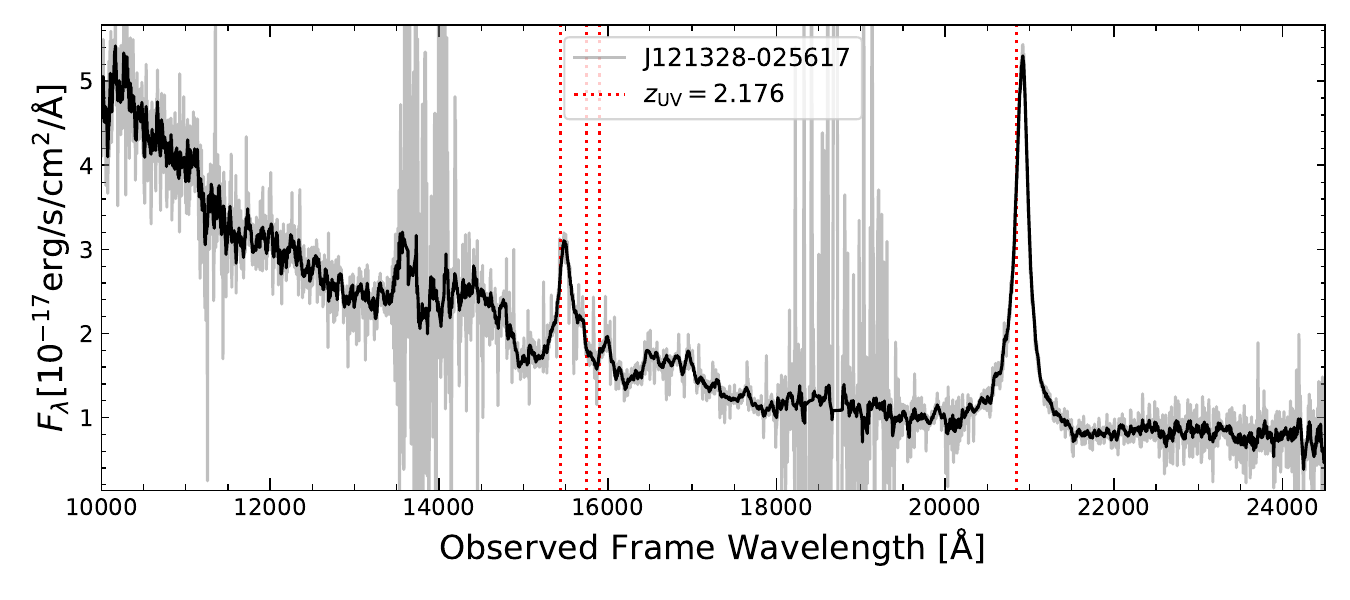}
    \includegraphics[clip=true, trim={10 0 10 10}, width=\columnwidth]{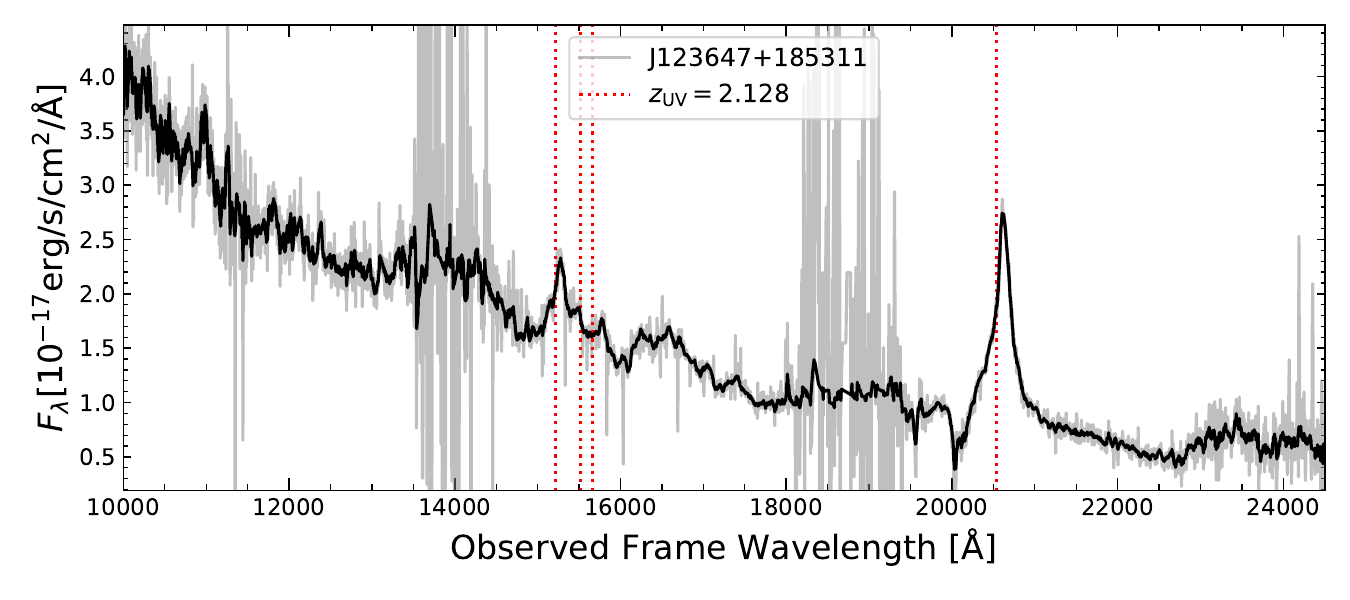}
    \caption{Magellan FIRE spectra for twelve quasars which are newly presented in this work (Section~\ref{sec:data:FIRE}), 
    {with a resolution of 50\kmps ($R=6000$).}
    Gray shows the spectra, with a 9-pixel inverse variance weighted smooth in black. The expected wavelengths of \hb, \oiii\ and \ha\ are marked in red, assuming the redshifts estimated from the SDSS rest-frame ultraviolet spectra.
    {The corresponding SDSS spectra are shown in Appendix~\ref{sec:app:SDSSspectra}.}
    }
    \label{fig:spectra}
\end{figure*}

Spectra were reduced using the standard set-up in \texttt{PypeIt} v1.11 \citep{2020JOSS....5.2308P, pypeit:zenodo}.
Targets were nodded along the slit in ABBA dither pattern and combined in AA-BB groups for sky subtraction. 
Wavelength calibration was performed using night-sky emission features.
Bright A0V stars  observed directly before or after each science target were  used for flux calibration. Spectra were then corrected for telluric absorption using a model grid.
We show the reduced \textit{JHK} spectra in Fig.~\ref{fig:spectra}: some residual telluric features can be observed around 20000-20150\,\AA, but \ha\ and \hb\ are clearly detected in each object. \oiii\ is clearly seen in e.g. J100711+053208, but in many objects there is strong \ion{Fe}{II} emission in the \textit{H} band and spectral modelling is required to deblend \hb, \oiii\ and \ion{Fe}{II}.

\subsection{Existing near-infrared spectra}
\label{sec:data:existingNIR}
To complement our targeted observations with Magellan/FIRE, we also cross-match our parent sample of SDSS quasars to various catalogues from the literature to build a large sample of objects with coverage of \civ\ and \oiii.
Starting from the SDSS sample described in Section~\ref{sec:data:SDSS}, we look for near-infrared data from the GNIRS `Distant Quasar Survey' \citep{2021ApJS..252...15M, 2023ApJ...950...95M}, the previous compilation of \citet{Coatman17, Coatman19}, the X-Shooter programme described by \citet{2019ApJ...876..105X, 2020MNRAS.495..305X}, and the SUPER survey \citep{2018A&A...620A..82C}.
For objects which have more than one near-infrared spectrum available, we keep only the spectrum with the highest signal-to-noise ratio in the 4800--5100\,\AA\ region which contains \hb\ and \oiii.

\subsubsection{GNIRS data from the DQS}

The Gemini/GNIRS `Distant Quasar Survey' \citep[DQS; PI: Shemmer;][]{2021ApJS..252...15M, 2023ApJ...950...95M} is a large and long programme to obtain high signal-to-noise ratio near-infrared spectra for a large sample of SDSS quasars at $1.5<z<3.5$.
Raw data were downloaded from the Gemini archive and reduced using \texttt{PypeIt}.
{190}
objects from the GNIRS-DQS match to our sample of SDSS spectra.

\subsubsection{Coatman et al. compilation}

The largest previous catalogue of  \oiii\ emission from near-infrared quasar spectra was described  by \citet{Coatman19}.
This compilation includes near-infrared spectra from 
WHT/LIRIS observations presented by \citet{Coatman16},
TRIPLESPEC and FIRE observations presented by \citet{2012ApJ...753..125S} and \citet{Shen16},
observations from the `Quasars probing Quasars' project \citep{2006ApJ...651...61H},
as well as programmes from P200/TRIPLESPEC, VLT/SINFONI and NTT/SOFI.
The construction of this data set is described in full by \citet{Coatman17}.
For the purposes of this investigation, we keep only spectra with median signal-to-noise ratio (per 69\,km/s pixel) $>3$ across the rest-frame 4800-5100\,\AA\ region.
{One-hundred}
spectra from this catalogue with coverage of \hb, \oiii\ and \ha\ are matched to our SDSS sample.

\subsubsection{X-Shooter data from Xu et al.}

\citet{2020MNRAS.495..305X} presented near-infrared observations of five BAL and two miniBAL quasars. These data were taken from a wider VLT/X-Shooter programme (PI: Benn) which observed a total of 20 quasars with rest-frame ultraviolet absorption features \citep{2019ApJ...876..105X}.
Ten of these X-Shooter sources match to our SDSS parent sample. 
For each source, we downloaded the Phase 3 data products from the ESO archive. Each exposure was corrected for telluric absorption using \texttt{molecfit} \citep{2015A&A...576A..78K, 2015A&A...576A..77S}. Where more than one exposure was present for a source, all such observations were combined to give the best signal-to-noise ratio spectrum for each quasar.

\subsubsection{SINFONI data from the SUPER survey}

The SUPER survey \citep{2018A&A...620A..82C, 2020A&A...642A.147K, 2020A&A...644A.175V} is a VLT/SINFONI large programme (PI: Mainieri) designed to study AGN feedback at so-called `cosmic noon', $z\approx 2$.
Targets were selected from X-ray surveys to have redshifts $2<z<2.5$. We downloaded the DR1 data products from the ESO archive, which consist of the combined, flux-calibrated data cubes for 20 Type-1 AGN presented by \citet{2020A&A...642A.147K}.
We extract 1d spectra from the SUPER data cubes using 0.6\,arcsec apertures to match the slit width of our FIRE observations, and estimate 1d noise arrays from the pixel-pixel variations away from the source in each data cube.
{We find that one}
SUPER source  with  SINFONI \textit{H+K} data covering \hb, \oiii\ and \ha, has SDSS data which meet the criteria described in Section~\ref{sec:data:SDSS}.

\section{Methods and Results}
\label{sec:methods}

\subsection{Spectral modelling procedure}

We use the open source code \textsc{fantasy}\footnote{\url{https://fantasy-agn.readthedocs.io}} to simultaneously model the \oiii, \ion{Fe}{II}, and Balmer emission lines in each observed-frame optical spectrum from our sample of quasars.
\textsc{fantasy} is a python code for simultaneous multi-component fitting of AGN spectra,
described by \citet{2020A&A...638A..13I, 2023ApJS..267...19I} and \citet{2022MNRAS.516.1624R}.
For our luminous ($10^{46}\lesssim L_\textrm{bol}\lesssim 10^{48}\,\ergps$) quasars, we do not include a host galaxy component.
The Balmer lines (\ha\ and \hb, and where available  \hg\ and \hd) are kinematically tied to have identical velocity profiles, with up to two broad and one narrow Gaussian components, although the relative normalisation of each line is free to vary.
A comprehensive set of \ion{Fe}{II} emission blends are included as described in \citet{2023ApJS..267...19I}.
We require coverage of \ha\ to ensure robust modelling of the red wing of \hb\ in objects with weak \oiii\ and strong \ion{Fe}{II} blended around 4900-5100\,\AA.

We fit each spectrum twice: once with no \oiii, and once with one broad and one narrow \oiii\ component.
The 4960 and 5008\,\AA\ lines are constrained to have identical kinematics with the amplitudes tied in a $1:3$ ratio.
The Bayesian Information Criterion (BIC) is calculated for each fit.
For quasars where the BIC is not improved by 10 or more when including \oiii, we flag the \oiii\ component as not robustly detected, and exclude the spectrum from our kinematic analysis.
Many of these spectra would be inferred to have extremely weak
(rest-frame) \oiii\ EW $<1$\,\AA. Two-hundred and twenty-eight objects from our sample of 313 quasars are judged to have robust \oiii\ detections: 57 BALs, 78 miniBALs and 93 non-BALs.
For these objects we measure $w_{80}$, the  velocity width containing 80 per cent of the 
{total}
5008\,\AA\ line flux, as is commonly used in the literature to quantify \oiii\ outflow signatures \citep{Zakamska14, Coatman19, 2020A&A...634A.116V}.

\subsection{Results}
\label{sec:results}

\begin{figure*}
    \centering
    \includegraphics[width=\textwidth]{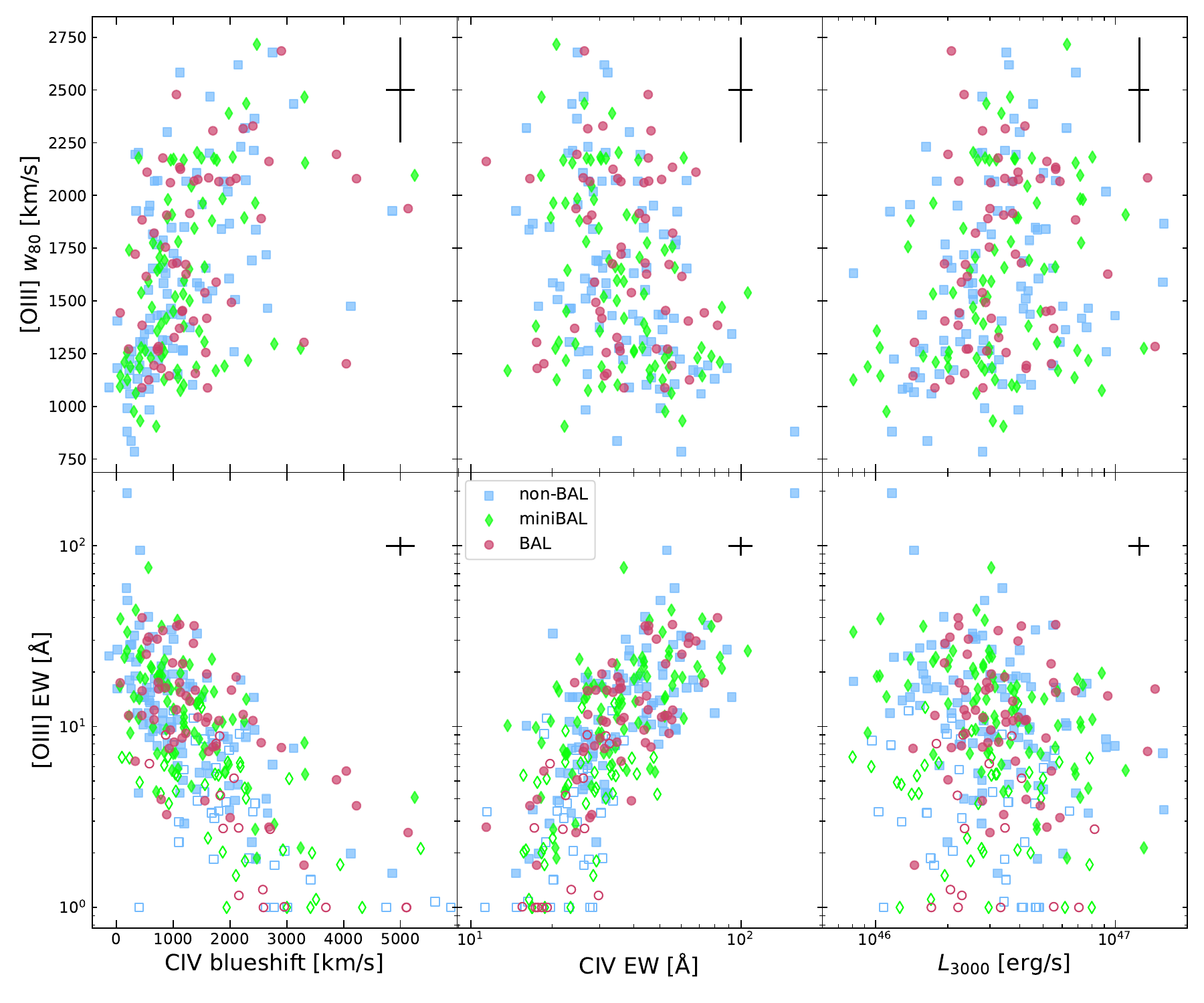}
    \caption{\oiii\ EW and $w_{80}$ as a function of \civ\ properties and $L_\textrm{3000}$.
    {The typical uncertainty associated with each individual data point is shown by the black crosses in the upper right of each panel.
    Open symbols indicate sources where $\Delta\mathrm{BIC} < 10$ when adding  \oiii\ to the spectral model; these \oiii\ lines are not considered to be robustly detected and so we do not attempt to derive velocity width information.
    Objects with weaker \oiii\ are plotted at 1\,\AA\ EW for display purposes only. }
     The \oiii\ strength and velocity width correlate with the \civ\ emission line blueshift and  EW, with the same trends observed in our samples of BAL, miniBAL and non-BAL quasars. 
     }
    \label{fig:EWs}
\end{figure*}

Our first observational result is the dependence  of the \oiii\ EW on the \civ\ emission-line blueshift and EW, as shown in the bottom panels of Fig.~\ref{fig:EWs}.
For objects with robust \oiii\ measurements, we show the $w_{80}$ velocity width in the top panels of Fig.~\ref{fig:EWs}.
Consistent with previous works \citep[e.g.][]{Coatman19}, we find that the \oiii\ properties are correlated with the ultraviolet \civ\ morphology.
Objects with larger \civ\ blueshifts have weaker \oiii\ EW 
(Pearson's $r=-0.42$, $p=10^{-14}$)
with broader \oiii\ $w_{80}$
($r=0.45$, $p=10^{-12}$).
Stronger \civ\ EW 
is correlated with
stronger \oiii\ EW 
($r=0.65$, $p=10^{-39}$)
and narrower \oiii\ $w_{80}$
($r=-0.34$, $p=10^{-7}$).
When we compare our BAL, miniBAL and non-BAL quasar samples, we see no difference in their \oiii\ emission, when the underlying dependence on the  \civ\ emission properties is taken into account.
In other words, we see no evidence for BAL or miniBAL quasars having different narrow line region properties when compared to their non-BAL counterparts.

We also see a weak trend with luminosity: objects with larger $L_{3000}$ are more likely to show smaller \oiii\ EW 
{($r=-0.20$, $p=0.00037$)}
and broader \oiii\ $w_{80}$
{($r=0.23$, $p=0.00048$),}
consistent with previous works \citep{Zakamska14, Coatman19, 2020A&A...634A.116V}.
However, the dynamic range in luminosity spanned by the majority of our sample is relatively small ($\approx$1\,dex), and the correlation observed between \oiii\ and \civ\ is not a secondary effect driven by an underlying correlation with the intrinsic quasar luminosity (such as the Baldwin effect).

\begin{figure*}
    \centering
    \includegraphics[width=\textwidth]{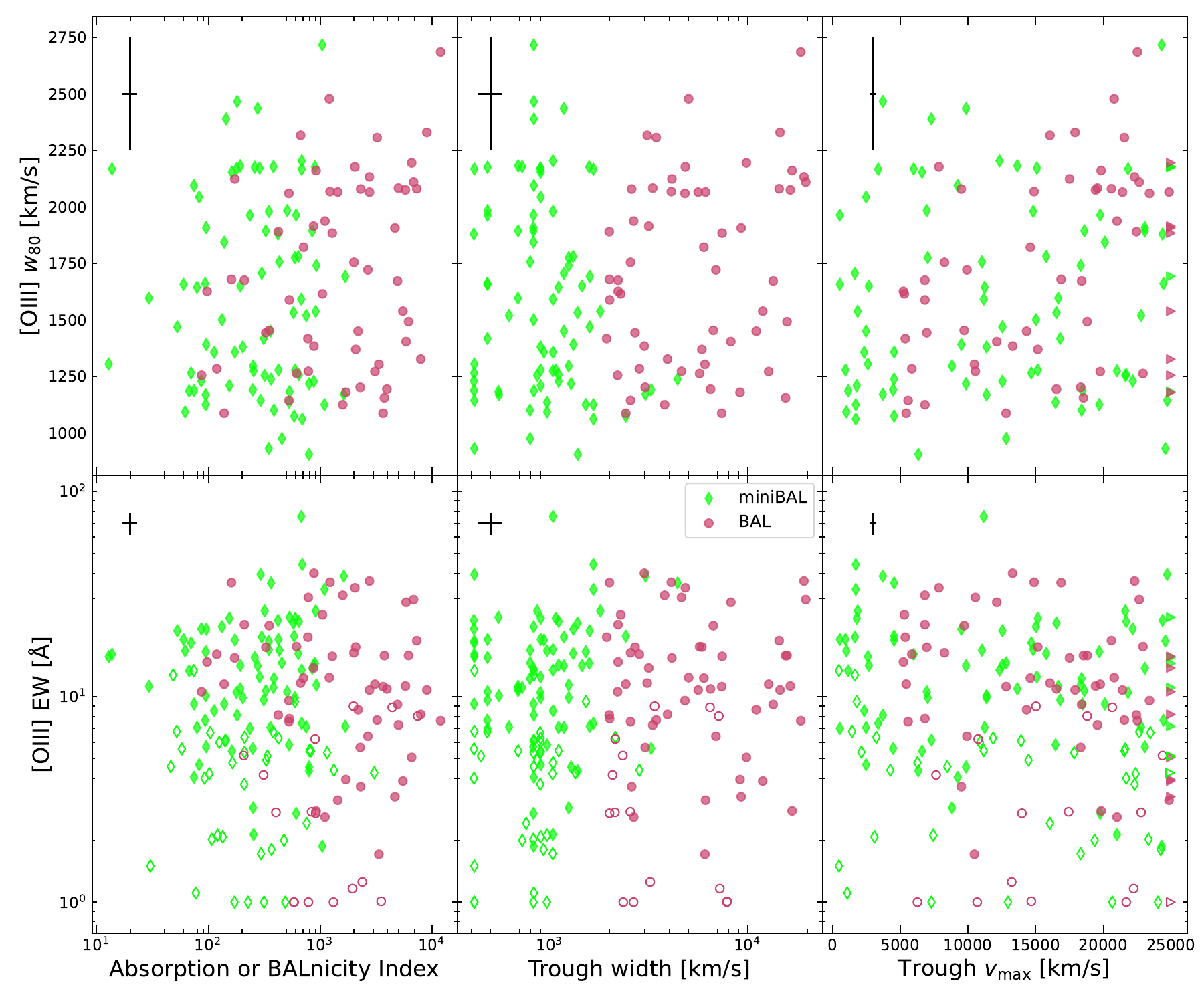}
    \caption{\oiii\ EW and velocity width as a function of \civ\  absorption properties.
    {The typical uncertainty associated with each individual data point is shown by the black crosses in the upper left of each panel.}
    As in Fig.~\ref{fig:EWs}, open symbols indicate objects where \oiii\ was not robustly detected in the near-infrared spectrum.
    For BALQSOs, we show the BI (Equation~\ref{eq:BI}) and the width and maximum velocity of the BAL trough.
    For miniBALs, we show the AI (Equation~\ref{eq:AI}) and the width and maximum velocity of the miniBAL trough.
    For objects with more than one trough, we plot the median trough width and the fastest maximum velocity. Our absorption-finding algorithm only searches up to 25\,000\,\kmps, to avoid confusion with \ion{Si}{IV} absorption, so the cluster of points on the right-hand edge of the right-hand panels 
    {(shown as triangles)}
    could be considered lower limits on $v_\textrm{max}$.
    }
    \label{fig:BAL_results}
\end{figure*}

For our BAL and miniBAL quasar samples,
we  test if the absorption trough properties are linked to the kinematics of the narrow line region traced by the  \oiii\ emission.
In Fig.~\ref{fig:BAL_results} we show the \oiii\ EW and $w_{80}$ as a function of absorption trough properties for both the miniBAL and the \textit{bona fide} BAL populations.
We show the \citet{1991ApJ...373...23W} BALnicity Index (BI)  and  \citet{2002ApJS..141..267H} Absorption Index (AI), the absorption trough width (or median width, for quasars with more than one trough), and the fastest outflow velocity of the absorption trough 
(i.e.\ the fastest velocity contributing to the BI or AI measurements in Equations~\ref{eq:BI} and \ref{eq:AI}), 
as these parameters have been seen to show the strongest correlations  with \civ\ emission line parameters \citep{Rankine20}. 
We find a mild but statistically significant correlation between the maximum  BAL trough velocity and the \oiii\ $w_{80}$ 
{($r=0.37$, $p=0.004$). }
However, if we express the maximum trough velocity in units of the \civ\ emission-line blueshift the correlation with \oiii\ is not present, suggesting that the BAL is tracing the high-velocity tail of the \civ\ emission-line outflow, $v_\textrm{max}\approx10$ times the median \civ\ emission blueshift.
None of the other BAL trough parameters computed by \citet{Rankine20}, such as the minimum trough velocity and velocity of the deepest part of the trough, is found to correlate with the \oiii\ measurements computed in this work.

\section{Discussion}
\label{sec:discuss}

The observational results presented in the previous section can be summarised as three key findings.
\textit{First,} larger \civ\ emission-line  blueshift is observed to correlate with weaker \civ\ EW, weaker \oiii\ EW and broader \oiii\ velocity width, confirming the correlations found in smaller samples by previous authors \citep{2004ApJ...617..171B, 2017FrASS...4...16M, 2018A&A...617A..81V,  Coatman19, 2020A&A...644A.175V, 2023A&A...669A..83D}.  
\textit{Second,} BAL and miniBAL quasars do not show significant differences in their \oiii\ emission  compared to their non-BAL counterparts.
\textit{And third,} while there is a mild correlation between the BAL outflow velocity and the \oiii\ velocity width, this is most likely driven by the fact that these parameters are both correlated with the  underlying \civ\ emission-line kinematics \citep{Coatman19, Rankine20}.

In this section, we first discuss these findings in the context of previous work, and then explore  possible interpretation.

\subsection{Comparison with previous BAL investigations}

We believe our investigation into the rest-frame optical \oiii\ properties of 73 BAL quasars represents the largest such sample available to date.
Recently \citet{2020MNRAS.495..305X} conducted a similar investigation, measuring the \oiii\ strength and kinematics in a smaller sample of five BAL and two miniBAL quasars. 
\citet{2020MNRAS.495..305X}  selected their sample to have high-ionization \ion{Si}{IV} BALs in addition to \civ\ troughs, allowing them to infer constraints on the density and radius of the absorbing gas.
They find that the electron number density derived for the absorbing material increases with decreasing \oiii\ EW.\footnote{Xu et al. report $L_{\oiii}/L_\textrm{bol}$, which is closely related to the EW.}
They also find that the measured velocity widths have similar sizes in the BAL troughs and \oiii\ emission lines, and suggest that this is consistent with the \oiii\ emission and BAL absorption being ``different manifestations of the same wind''.
Our \oiii\ velocity widths span a similar range to those found by \citet{2020MNRAS.495..305X}, with $w_{80}$ in the range 1000-2500\,\kmps\ and $w_{90}$ in the range 1000-3000\,\kmps.
However, our BAL trough widths span a larger range, up to more than 20,000\,\kmps, and with our larger sample we do not find a significant correlation between the BAL trough widths and the \oiii\ velocity widths. 
Unlike \citet{2020MNRAS.495..305X}, we therefore do not believe that the \oiii\ emission and BAL absorption need to be tracing the same gas in each object, that is to say outflowing gas with the same density, ionization parameter and location. 

As this paper was going through the peer review process, 
\citet{2024ApJ...968...77A}
was posted on the arXiv. These authors investigated 65 BAL quasars from the GNIRS-DQS and found no significant differences in their rest-frame optical spectra compared to a control sample of non-BAL quasars. Our results are in agreement with \citet{2024ApJ...968...77A}, although we note that our larger sample size and new Magellan/FIRE data together also allow us to investigate the \oiii\ emission in a statistical sample of
 BALs with more extreme \civ\ blueshifts $\gtrsim2500\kmps$.

In this work we have focused on high-ionization BAL quasars, which do not show absorption troughs from low-ionization lines such as \mgii\,$\lambda$2800 and \ion{Al}{III}\,$\lambda$1860.
Quasars with such low-ionization BAL troughs (`LoBALs') can be identified at lower redshifts when \mgii\ or \ion{Al}{III} is  present in the SDSS observed-frame optical spectrum \citep{1993ApJ...413...95V}.
\citet{2017ApJ...848..104S} presented near-infrared observations of 22 LoBALs at $1.3<z<2.5$, finding no enhancement in \oiii\ outflow signature compared to the wider non-BAL population.
\citet{2017MNRAS.467.2571M} also found no difference in distribution of \oiii\ EWs in 58  LoBALs at $0.35<z<0.83$, when compared to wider SDSS quasar population at the same redshift.
Our results are consistent with these works to the extent that neither our high-ionization BAL quasars nor their low-ionization BAL quasars show any significant difference from the non-BAL population in terms of their optical \oiii\ emission, suggesting that the narrow line region properties are not significantly impacted by the presence of a broad absorption feature in the rest-frame ultraviolet.

\subsection{Implications for quasar winds and feedback}

This investigation was motivated (in part) by the idea that if BAL and non-BAL quasars are members of two distinct AGN populations with different disc-winds, driving ionized gas outflows with differing powers, then we might expect BALs and non-BALs to show differences in their narrow line region (as traced by the \oiii\ strength and kinematics).
However, this is true only if the length of the BAL phase is comparable to the time taken to affect the narrow line region.
Estimates of quasar lifetimes are typically of the order of 1-10\,Myr \citep{2021MNRAS.505..649K}, which together with the significant ($\sim$10-40 per cent) observed BAL fraction \citep{2011MNRAS.410..860A} means that
if the BAL phenomenon was a single evolutionary phase then it would typically last at least $\sim0.1$\,Myr. In this time a BAL outflow travelling at 10\,000\,\kmps\ would travel $\sim$1\,kpc, meaning that we would expect it to reach the scales which we are probing with \oiii.
In Section~\ref{sec:results} we found that the BAL, miniBAL and non-BAL quasar populations show no significant differences in their narrow line region \oiii\ properties. Our results therefore suggest that the BAL phenomenon is not a distinct, long-lived evolutionary phase in the cycle of  SMBH growth, at least within the context of luminous quasar activity.

One possible interpretation could instead be that BAL outflows don’t do anything to affect the \oiii-emitting gas in luminous quasars -- either suggesting that the kinetic power contained in BAL outflows is negligible, or that they are entrained in a particular geometry which means that they are directed away from the narrow line region gas.
This would imply either that BAL outflows are not important for quasar feedback, which is unlikely given that photoionization models of BAL troughs suggest that at least some BALs carry significant kinetic power \citep{2018ApJ...857...60A,2019ApJ...876..105X}, or that the \oiii-emitting region is not tracing the impact of quasar-mode outflows on the interstellar medium of the AGN's  host galaxy.

On the other hand, if we assume that BAL outflows should affect the narrow line region, then the similarity seen in the BAL and non-BAL quasars would suggest that the 
{observation of a BAL trough}
 is an intermittent, but recurring phenomenon during the luminous quasar phase of the SMBH growth cycle. There is a growing body of work suggesting that BAL troughs can vary on relatively short timescales, which would support this hypothesis \citep{2008MNRAS.391L..39H,2008ApJ...675..985G,2010ApJ...713..220G,2011MNRAS.413..908C,2013MNRAS.429.1872C,2012ApJ...757..114F,2013ApJ...777..168F,2015ApJ...806..111G,2017MNRAS.469.3163M,2018A&A...616A.114D,2018ApJ...862...22R,2019ApJ...872...21H,2021MNRAS.504.3187M,2022A&A...668A..87V,2023MNRAS.522.6374A}.
In other words, our results would be consistent with a scenario where BAL troughs are an intermittent tracer of a persistent quasar outflow (which could be the wind traced by the \civ\ emission blueshift), where a part of the outflow along our line-of-sight becomes optically thick for short periods of time when a broad absorption trough is observed.
This scenario would explain the correlations observed between the \civ\  and \oiii\ outflow signatures: luminous quasars which drive  winds on nuclear scales are also able to drive outflows on much larger scales, with stochastic parts of the wind sometimes producing absorption features which don't correlate directly with the properties of the narrow line region outflow. The general properties of such nuclear winds would then be governed by the shape and strength of the ionizing quasar continuum, which is in turn set by the SMBH mass and accretion rate \citep{Temple23}.
This scenario would also explain the fact that BAL and non-BAL quasars have no significant difference in their underlying \civ\ emission  \citep{Rankine20} or their sublimation-temperature dust emission (section 4.3 of \citealt{2021MNRAS.501.3061T},
{see also \citealt{2023A&A...671A..34S}),}
but a further element would be required to explain the differences in  radio properties of the BAL and non-BAL populations observed by \citet{2022MNRAS.515.5159P,2024MNRAS.529.1995P}, especially as the radio emission in AGN is often found to be linked with the \oiii\ properties \citep[e.g.][]{2021MNRAS.503.1780J}.

\section{Conclusions}
\label{sec:conclude}

We have measured the rest-frame optical \oiii\,$\lambda\lambda$4960,5008 emission in a sample of 73 BAL, 115 miniBAL and 125 non-BAL
quasars at $1.56<z<2.6$ with high-quality near-infrared spectroscopic data.
Our key observational results are:
\begin{enumerate}
    \item The properties of \oiii\,$\lambda$5008  and \civ\,$\lambda$1550 emission are connected: larger \civ\ emission-line  blueshift and weaker \civ\ EW correlate with weaker \oiii\ EW and broader \oiii\ velocity structure. 
    These correlations are not driven solely by changes in the 3000\,\AA\ continuum luminosity.
    \item BAL, miniBAL and non-BAL quasars show no significant differences in their \oiii\ emission: all three sub-samples show the same correlations described in (i).
    \item In BALQSOs, the maximum absorption trough velocity  shows a weak correlation with the \oiii\ velocity width, but this is fully explained by the dependence of both quantities on the \civ\ emission-line blueshift. 
    When the BAL outflow velocities are normalised by the \civ\ blueshift, we find no correlations between the \oiii\ emission kinematics and the \civ\ BAL trough parameters.
\end{enumerate}

Our results disfavour a scenario where (high-ionization) BALQSOs are a special, long-lived evolutionary phase in which more powerful (cf. non-BAL QSOs) winds are able to propagate into the interstellar medium of their host galaxies, and either clear out or alter the kinematics of this medium.
Instead our results are consistent with a scenario in which BAL troughs are an intermittent, stochastic phenomenon which all luminous quasars with persistent outflowing winds (traced by both \oiii\ and \civ\ emission) are likely to undergo.
This is consistent with observations of BAL trough variability on time-scales of only months to years,
{and supports the scenario favoured by \citet{Rankine20} where BALs and non-BALs are members of the same underlying quasar population.}

\section*{Acknowledgements}

MJT acknowledges funding from a FONDECYT Postdoctoral fellowship (3220516) and from STFC (ST/X001075/1). 
ALR acknowledges support from UKRI (MR/T020989/1).
MB and JM acknowledge funding from the Royal Society via University Research Fellowships (URF/R/221003 and URF/R1/221062).
CR acknowledges support from Fondecyt Regular grant 1230345 and ANID BASAL project FB210003.

This research made use of \textsc{Astropy}\footnote{\url{http://www.astropy.org}}, a community-developed core Python package and an ecosystem of tools and resources for astronomy \citep{astropy:2013, astropy:2018, astropy:2022}, 
 \textsc{matplotlib} \citep{Hunter:2007}, \textsc{numpy} \citep{numpy},
 \textsc{fantasy}\footnote{\url{https://fantasy-agn.readthedocs.io}} 
 \citep{2020A&A...638A..13I, 2022MNRAS.516.1624R, 2023ApJS..267...19I} and \textsc{PypeIt},\footnote{\url{https://pypeit.readthedocs.io/en/latest/}}
a Python package for semi-automated reduction of astronomical slit-based spectroscopy
\citep{2020JOSS....5.2308P}.
MJT thanks Dalia Baron and Christina Eilers for advice in using \textsc{PypeIt}, Xinfeng Xu for advice on the X-Shooter sample, Vincenzo Mainieri for advice on the SUPER data, Brandon Matthews for advice on GNIRS-DQS, and Bartolomeo Trefoloni for access to data from \citet{2023A&A...677A.111T}.
We thank the anonymous referee for a thorough report, which helped to improve the manuscript. MJT would also like to thank Kate Grier for a useful discussion on BAL variability.

This paper includes new data gathered with the 6.5 meter Magellan Telescopes located at Las Campanas Observatory, Chile through CNTAC programme CN2020B-4 (PI: Temple).
These observations were originally scheduled for July 2020; we thank CNTAC for carrying-over the programme to January 2022. 

This paper uses data from the SUPER survey \citep{2018A&A...620A..82C}
based on data products created from observations collected at the European Organisation for Astronomical Research in the Southern Hemisphere under ESO programme 196.A-0377.
This paper also uses X-Shooter data from ESO programmes 087.B-0229,
090.B-0424,
091.B-0324 and
092.B-0267.
This research has made use of the services of the ESO Science Archive Facility.

This paper uses data from GNIRS-DQS \citep{2021ApJS..252...15M, 2023ApJ...950...95M}, based on observations obtained at the international Gemini Observatory with program IDs GN-2017B-LP-16, GN-2018A-LP-16, GN-2018B-LP-16, GN-2019A-LP-16, GN-2019B-LP-16, GN-2020A-LP-16, and GN-2020B-LP-16.
Gemini is a program of NSF’s NOIRLab, which is managed by the Association of Universities for Research in Astronomy (AURA) under a cooperative agreement with the National Science Foundation on behalf of the Gemini Observatory partnership: the National Science Foundation (United States), National Research Council (Canada), Agencia Nacional de Investigaci\'{o}n y Desarrollo (Chile), Ministerio de Ciencia, Tecnolog\'{i}a e Innovaci\'{o}n (Argentina), Minist\'{e}rio da Ci\^{e}ncia, Tecnologia, Inova\c{c}\~{o}es e Comunica\c{c}\~{o}es (Brazil), and Korea Astronomy and Space Science Institute (Republic of Korea).
This work was enabled by observations made from the Gemini North telescope, located within the Maunakea Science Reserve and adjacent to the summit of Maunakea. We are grateful for the privilege of observing the Universe from a place that is unique in both its astronomical quality and its cultural significance.

Funding for the Sloan Digital Sky Survey IV has been provided by the Alfred P. Sloan Foundation, the U.S. Department of Energy Office of Science, and the Participating Institutions. SDSS-IV acknowledges
support and resources from the Center for High-Performance Computing at
the University of Utah. The SDSS web site is www.sdss.org.


\section*{Data Availability}

The optical spectroscopic data underlying this article are available from SDSS.\footnote{\url{https://www.sdss4.org/dr17/}}
New Magellan/FIRE infrared spectra will be made available via CDS.
 A FITS file with the emission-line measurements used in this work is included as supplemental data. The file contains a table with 313 rows, one for each quasar, and columns as described in Table~\ref{tab:2}.

\begin{table*}
    \centering
    \caption{Description of columns for the supplemental data file available at MNRAS online.}
    \label{tab:2}
    \begin{tabular}{llrr}
        \hline
        Column name  & Description & Units & Example entry \\
        \hline
        \texttt{Name} & SDSS Name     &   --  &   J000730.94-095831.9 \\
        \texttt{RA} & Right Ascension & decimal  degrees   &   1.8789376910543467 \\
        \texttt{Dec} & Declination  &  decimal  degrees  &  -9.975545336959057 \\
        \texttt{z} & Redshift    &    --         & 2.224 \\
        \texttt{L3000} & Logarithm of 3000\,\AA\ luminosity & dex(\ergps) & 46.47 \\
        \texttt{Classification} & [BAL, miniBAL, non-BAL] & -- & BAL \\
        \texttt{CIV\_EW} & \civ\ EW   &   \AA ngstr\"om  &   35.99 \\
        \texttt{CIV\_blue} & \civ\ blueshift & \kmps\  & 861.4 \\
        \texttt{BI\_BI} & BALnicity index (Eq.~\ref{eq:BI}) & \kmps\  & 2003.4 \\
        \texttt{BI\_VMAX} & BAL trough maximum velocity & \kmps\  & 8271.8 \\
        \texttt{BI\_WIDTH} & BAL trough median width   &  \kmps\  & 2553.1 \\
        \texttt{AI\_AI} & Absorption index  (Eq.~\ref{eq:AI})  &  \kmps\  &   3348.6 \\
        \texttt{AI\_VMAX} & AI trough maximum velocity &  \kmps\  &   16815.9 \\
        \texttt{AI\_WIDTH} & AI trough median width   &  \kmps\   &   1311.5 \\
        \texttt{O3\_EW} & \oiii\ EW   &   \AA ngstr\"om   &   16.38 \\
        \texttt{O3\_w80} & \oiii\ 80 per cent velocity width  & \kmps\  & 1754.8 \\
        \texttt{O3\_w90} & \oiii\ 90 per cent velocity width  & \kmps\  & 2254.1 \\
        \hline
    \end{tabular}
\end{table*}



\bibliographystyle{mnras}
\bibliography{paper_refs} 




\appendix

\section{SDSS data}
\label{sec:app:SDSSspectra}
For each of the twelve quasars with new Magellan/FIRE spectra presented in Section~\ref{sec:data:FIRE}  (Fig.~\ref{fig:spectra}),  the SDSS rest-frame ultraviolet spectra and corresponding ICA reconstructions are shown in Fig.~\ref{fig:sdss_spectra}. 

\begin{figure*}
    \centering
    \includegraphics[width=\linewidth]{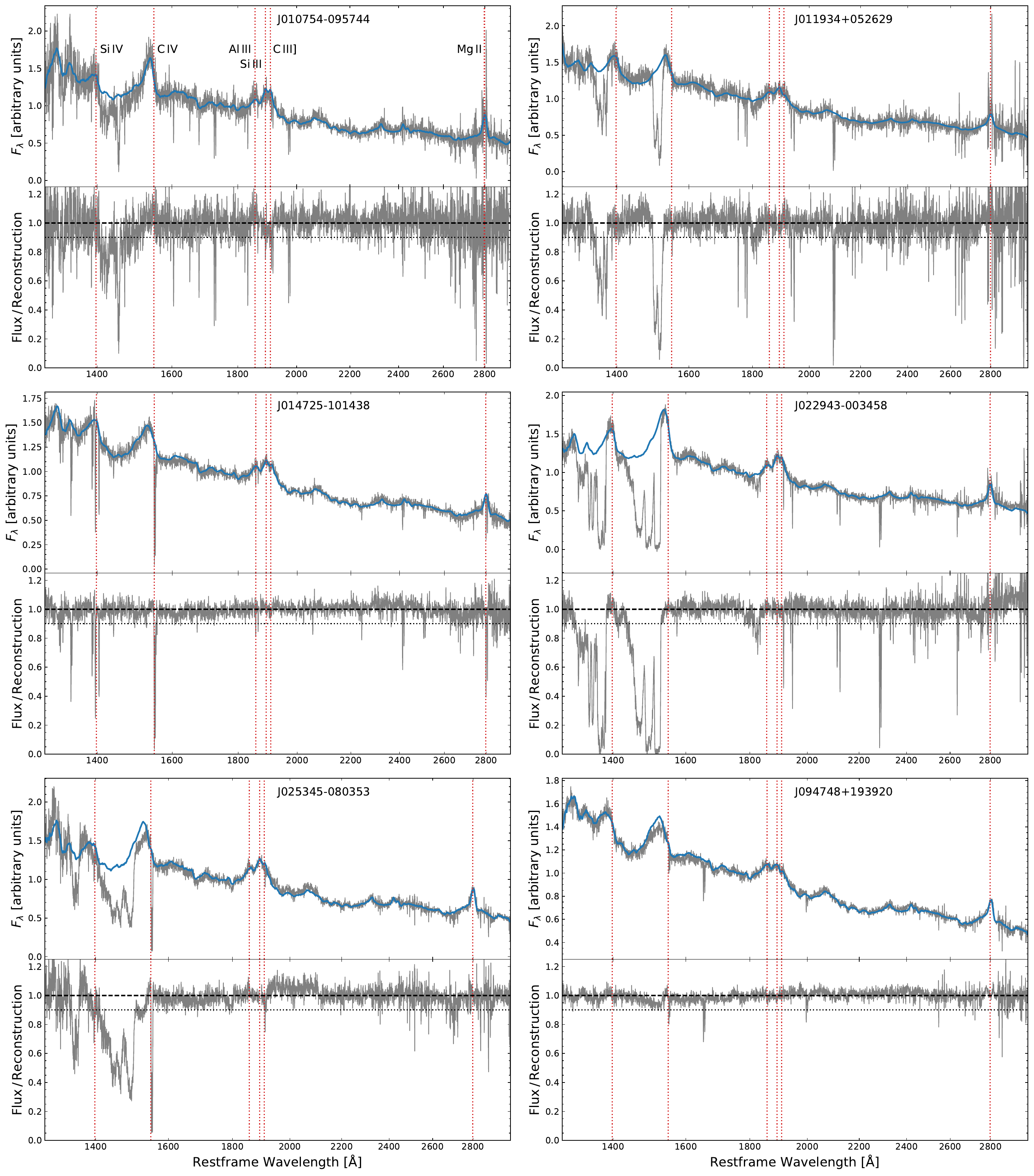}
    \caption{
    Examples of the rest-frame ultraviolet measurements used in this work. Red vertical lines show the rest-frame wavelengths of \ion{Si}{IV}, \civ, \ion{Al}{III}, \ion{Si}{III}], \ion{C}{III}], and \mgii.
    Top panels show the SDSS spectra in gray and corresponding ICA spectral reconstruction in blue. Absorption features are iteratively masked when modelling the observed data, allowing the ICA to reconstruct the unabsorped pixels.
    Lower panels show the ratio of the observed and reconstructed flux, with the dotted horizontal line showing the 0.9 threshold used to detect absorption features.
    BAL quasars are required to have at least 2000\,\kmps\ contiguous absorption (i.e. flux/reconstruction $<0.9$) starting from at least 3000\,\kmps\ bluewards of \civ\ (Eq.~\ref{eq:BI}).
    Objects with slower and/or narrower absorption features broader than 450\,\kmps\ are classed as `miniBALs' (Eq.~\ref{eq:AI}), otherwise the quasar is considered a `non-BAL'.
    }
    \label{fig:sdss_spectra}
\end{figure*}
\begin{figure*}
    \centering
    \includegraphics[width=\linewidth]{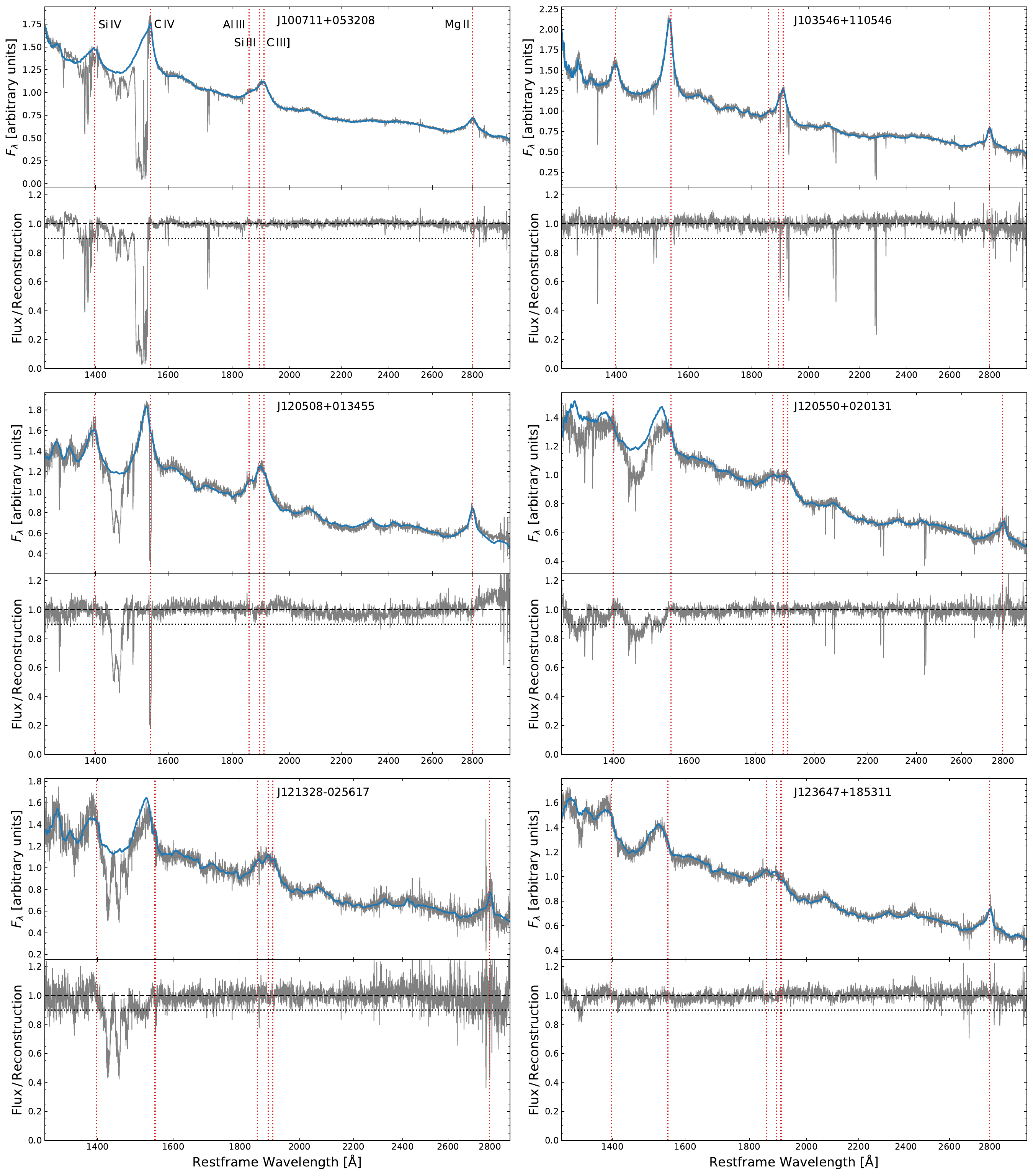}
    \contcaption{}
\end{figure*}


\bsp	
\label{lastpage}
\end{document}